# Dual scale Residual-Network for turbulent flow sub grid scale resolving: A prior analysis

Omar Sallam*,[a], Mirjam Fürth[a]

[a]Department of Ocean Engineering, Texas A & M University, College Station, TX, 77843, USA

ABSTRACT

This paper introduces generative Residual Networks (ResNet) as a surrogate Machine Learning (ML) tool for Large Eddy Simulation (LES) Sub Grid Scale (SGS) resolving. The study investigates the impact of incorporating Dual Scale Residual Blocks (DS-RB) within the ResNet architecture. Two LES SGS resolving models are proposed and tested for prior analysis test cases: a super-resolution model (SR-ResNet) and a SGS stress tensor inference model (SGS-ResNet). The SR-ResNet model task is to upscale LES solutions from coarse to finer grids by inferring unresolved SGS velocity fluctuations, exhibiting success in preserving high-frequency velocity fluctuation information, and aligning with higher-resolution LES solutions' energy spectrum. Furthermore, employing DS-RB enhances prediction accuracy and precision of high-frequency velocity fields compared to Single Scale Residual Blocks (SS-RB), evident in both spatial and spectral domains. The SR-ResNet model is tested and trained on filtered/downsampled 2-D LES planar jet injection problems at two Reynolds numbers, two jet configurations, and two upscale ratios. In the case of SGS stress tensor inference, both SS-RB and DS-RB exhibit higher prediction accuracy over the Smagorinsky model with reference to the true DNS SGS stress tensor, with DS-RB-based SGS-ResNet showing stronger statistical alignment with DNS data. The SGS-ResNet model is tested on a filtered/downsampled 2-D DNS isotropic homogenous decay turbulence problem. The adoption of DS-RB incurs notable increases in network size, training time, and forward inference time, with the network size expanding by over tenfold, and training and forward inference times increasing by approximately 0.5 and 3 times, respectively.

## 1. Introduction

Turbulent flow is a multiscale chaotic phenomenon, and achieving High-Resolution (HR) numerical simulation of such a flow is valuable for understanding the dynamics, energy balance, and statistics of the interactions between all scales: large, intermediate, and small ones [1].

Direct Numerical Simulation (DNS) [2] can resolve all scales in turbulent flows in high-resolution spatiotemporal domains by solving the Continuity equation 1 and Momentum equation 2 with no prior filtration or averaging. The numerical solution can be implemented in the spatial or the spectral domain [2]. In DNS, the grid size is smaller than the Kolmogorov length scale [3], and the energy contained in the small scales dissipates by the viscous friction. To achieve the high-resolution scale resolving in DNS, the grid size for 2-D and 3-D flows are in order of $L_0 Re^{-6/4}$ and $L_0 Re^{-9/4}$, respectively. $L_0$ is the integral length scale and $Re$ is Reynolds number [3]. The grid size requirements make the DNS computationally infeasible for practical flow problems, especially at high Reynolds numbers.

$$\frac{\partial u_i}{\partial x_i} = 0 \qquad (1)$$

$$\frac{\partial u_i}{\partial t} + u_j \frac{\partial u_i}{\partial x_j} = -\frac{1}{\rho}\frac{\partial p}{\partial x_i} + \nu \frac{\partial^2 u_i}{\partial x_j \partial x_j} + f_i \qquad (2)$$

Due to the infeasibility of DNS to resolve all turbulent scales for practical flow applications, turbulent flow models have been developed for decades, and are widely, used such as Reynolds Averaged Navier Stoke's (RANS) [4] and Large Eddy Simulation (LES) [5]. These turbulent models resolve the mean (RANS) or the filtered (LES) large-scale velocity/vorticity fields on the numerical grids, while the small-scale fluctuations effects were being modeled to overcome the closure problem of the unresolved small-scale fluctuations stresses. Until recently, LES was also computationally infeasible for practical engineering flow problems due to the grid size requirements. However, with the improvement of computational power, LES has become a feasible tool for studying the large/small turbulent scale interactions for real engineering applications at small and moderate ranges of Reynolds number.

*Corresponding author
ORCID(s):

Large Eddy Simulation was initially proposed by Joseph Smagorinsky in 1963 to simulate atmospheric air currents [5]. In LES, the instantaneous velocity field $u_i$ is decomposed into a resolved filtered large-scale resolved field $\overline{u_i}$ and unresolved Sub Grid Scale (SGS) fluctuation field $u'_i$, such that $u_i = \overline{u_i} + u'_i$, where $(\overline{\cdot})$ is the filter operator. By applying the filter operator $(\overline{\cdot})$ to the Continuity equation 1 and the Momentum equation 2, the spatially/temporally filtered velocity and pressure fields can be expressed by the filtered Continuity equation 3 and the filtered Momentum equation 4. In the filtered Momentum equation 4, a new term arises $\frac{\partial \tau_{ij}}{\partial x_j}$, where $\tau_{ij}$ is the Sub Grid Scale (SGS) stress tensor. From equation 5, the SGS stress tensor $\tau_{ij}$ is dependent on the SGS velocity fields $u'_i$, such that $\overline{u_i u_j} = \overline{(\overline{u_i} + u'_i)(\overline{u_j} + u'_j)}$, which is unresolved on the numerical grid, this is known as the turbulence closure problem [4]. To overcome the turbulence closure problem, LES turbulence models were being developed for decades to account for the SGS stress tensor $\tau_{ij}$ in the filtered momentum equation 4. Computation of the SGS stress tensor $\tau_{ij}$ can be implemented by various methods, such as implicit grid techniques or explicit turbulence models [4]. The first LES turbulence model, the Smagorinsky SGS model, is the most commonly used one in commercial/opensource CFD codes [5]). In the Smagorinsky model, the SGS stress tensor $\tau_{ij}$ can be computed from the resolved velocity field as shown in equation 6, where $\overline{S_{ij}}$ is the resolved strain rate tensor described by equation 7, $\Delta$ is the filter width, $C_s$ is the Smagorinsky coefficient, and $|\overline{S}|$ is the strain rate magnitude described by equation 8.

$$\frac{\partial \overline{u_i}}{\partial x_i} = 0 \qquad (3)$$

$$\frac{\partial \overline{u_i}}{\partial t} + \overline{u_j}\frac{\partial \overline{u_i}}{\partial x_j} = -\frac{1}{\rho}\frac{\partial \overline{p}}{\partial x_i} + \nu\frac{\partial^2 \overline{u_i}}{\partial x_j \partial x_j} + f_i - \frac{\partial \tau_{ij}}{\partial x_j} \qquad (4)$$

$$\tau_{ij} = \overline{u_i}\,\overline{u_j} - \overline{u_i u_j} \qquad (5)$$

$$\tau_{ij} - \frac{1}{3}\tau_{kk}\delta_{ij} = -2(C_s\Delta)^2 |\overline{S}|\,\overline{S_{ij}} \qquad (6)$$

$$\overline{S_{ij}} = \frac{1}{2}(\frac{\partial \overline{u_i}}{\partial x_j} + \frac{\partial \overline{u_j}}{\partial x_i}) \qquad (7)$$

$$|\overline{S}| = (2\overline{S_{ij}}\,\overline{S_{ij}})^{\frac{1}{2}} \qquad (8)$$

With the rapid development of Graphical Processing Units (GPUs) and Machine Learning architectures, ML turbulence closure models have become one of the top research areas in the turbulence modeling field, in the past decade [6]. With ML, Surrogate turbulence models can be trained to compute Reynolds stresses for RANS, the Sub Grid Scale stresses (SGS) for LES problems or they can upscale the Low-Resolution (LR) CFD simulations to high-resolution ones [7] while respecting the underlying physics (Turbulence super-resolution). Many of these physics-informed surrogate ML turbulence models are Multi-Layer Perceptron (MLP) networks or Convolutional Neural Networks (CNN) based. For 3-D applications CNNs are more computationally effective regarding GPU memory utilization, however, MLP networks allow dealing with agnostic mesh such as unstructured grids with no prepossessing such as conformal mapping that transforms the unstructured grid to a structured one to adapt the CNN architectures [8]. For physics-informed MLP networks, a supervised learning approach is commonly used. However, for the CNN-based models, both supervised and unsupervised learning are being developed [9].

Since turbulent flows are broad-spectrum problems where high and low spatial/temporal frequencies exist, extra care is required when designing MLP or CNN for LES super-resolution or closure modeling to avoid spectral bias phenomenon [10]. Spectral bias is the tendency of the network to learn low-frequency features faster than high-frequency features, and learning the high-frequency features is not guaranteed for both MLP [10] and CNN [11] architectures. To avoid the spectral bias in any general ML regression problem, modification has to be implemented to the model architecture, such as using Fourier-Features in MLP [12] or using multiscale convolution kernels in CNN [13], [14], [15].

Using a MLP Physics Informed Neural Network (PINN) a super-resolution is implemented for turbulent flow wakes downstream of a cylinder array in a 2-D LES flow [16]. The architecture does not require any prior training dataset as the network upscales the available sparse measurements to the high-resolution solution by satisfying the governing Partial Differential Equations (PDEs) such as Momentum and Continuity equations. However, this network cannot be used as a black box for other problems and the network needs to be fully trained or partially trained (Transfer learning) when applied to a different problem. They used the Fourier features mapping [12] that showed a dramatic improvement in learning and predicting the high-frequency spatial flow fluctuations mitigating the spectral bias phenomenon.

With a supervised fully connected perceptron network learning approach, Maulik and San [17] proposed an LES super resolution model using a shallow single-layer neural network. The input to the single-layer network is the filtered vorticity fields on a coarse square grid, and the output is the fine-resolved vorticity field. They trained and tested the model for three cases, 2-D turbulence homogeneous isotropic decay turbulence, 3-D Kolmogorov turbulence, and 3-D compressible stratified turbulence.

For CNN based SGS scale resolving models, it is common to use the architectures for general image transformation, segmentation, recognition, classification, denoising, deblurring, or super-resolution applications [18], [19]; these CNN architectures include but are not limited to Pix2Pix [20], UNet [21], Generative Residual Network (ResNet) [22], and Generative Adversarial Networks (GAN) [23], [24]. All these CNN architectures employ the skip connections [22] that dramatically mitigate the vanishing gradient problem [25] and the network degradation problem [26], allowing the design of much deeper CNNs.

Using a Convolutional Neural Network (CNN) based on a Residual Network (ResNet), a 3-D volumetric super-resolution of turbulent flow is implemented using a single convolution kernel with size [3 × 3] in the residual block [27]. The model was trained and tested for both prior and posterior problems. Prior testing was conducted on Filtered DNS (FDNS) data from homogeneous isotropic turbulence problems, where the effect of the filtering kernel on super-resolution accuracy was examined. It was found that the Gaussian Kernel yielded better results compared to the sharp spectral filter. For posterior testing, the super-resolution model was tested on an LES of homogeneous isotropic turbulence problems, showing promising results in upscaling the LES simulation to statistically align with the DNS solution

Physics Informed Super-Resolution Generative Adversarial Network (PI-SR-GAN) is developed to upscale the velocity and pressure fields from low-resolution to high-resolution [28]. The SR-GAN network includes a generator and discrimination, the generator task is the upscaling process, while the discriminator task is to judge if the generated fields are physically accepted or not. Both the generator and discriminator use [3×3] single convolution kernel residual blocks. They implemented prior testing on a Filtered DNS (FDNS) dataset of a 3-D incompressible forced isotropic homogeneous turbulence problem.

Moreover, Fukami *et al* [29] used a Multi-Scale Convolution Neural Network (MS-CNN) for a spatiotemporal turbulent flow super-resolution task, they trained and tested the model on a downsampled filtered DNS of 2-D laminar vortex shedding of cylinder flow, a 2-D decaying homogeneous isotropic turbulence problem, and 3-D channel flow datasets. For the laminar cylinder wake case, their model was able to reconstruct a high-resolution temporal evolution of the velocity fields from low-resolution inputs of the first and last frames. For the turbulent 2-D decaying homogeneous isotropic turbulence problem, the model was highly affected by the temporal range of the training data that needed to be tested on shorter time ranges for reasonable predictions. In the 3-D channel flow case, the MS-CNN model was only able to reconstruct the low-frequency features of the DNS reference data and failed to recover the high-frequency information.

Employing the unsupervised learning approach, Kim *et al* [9] employed the Cycle-GAN [30] for an unsupervised turbulent flow super-resolution task. With unsupervised learning, the CNN can be trained when pairs of low-resolution and high-resolution do not exist such as when the input low-resolution training dataset is LES and the output high-resolution dataset is a DNS. The unsupervised Cycle-GAN architecture is tested for a 3-D channel flow problem where LES is the input training dataset and DNS solution is the output dataset. The architecture successfully recovered the DNS statistics from the LES inputs at small upscale ratios and was unstable at the high upscale ratios.

Similar to the super-resolution task, MLP and CNN are being used for the SGS stress tensor $\tau_{ij}$ inference. Maulil *et al* [31] used a Multi-Layer Perceptron (MLP) Neural Network to predict the SGS stresses for LES, they tested the model on a filtered DNS two-dimensional decaying turbulence for prior and posterior tests. The input fields to their network are the vorticity, strain rate, stream function, and vorticity gradient.

Both MLP and CNN are used for the SGS stress tensor inference for a 2-D homogeneous isotropic turbulence decay problem [32]. They studied the effect of the input features on the accuracy of SGS stress tensor prediction. For input features that only include the filtered velocity components, the MLP failed to reproduce the SGS stress tensor unless neighboring stencil mapping was implemented, on the other hand, CNN was able to do so. By increasing the input features to include the velocity gradient and Laplacian, both MLP and CNN were able to reproduce the SGS stress tensor, however, the CNN architecture was more accurate. For a posterior analysis, the CNN architecture was able to infer the SGS stress tensor eight times faster than the dynamics Smagorinsky model.

Table 1
Literature review and present work contribution summary.

| Author/s | Application | NN | CNN | Supervised | Unsupervised | Prior | Postrior |
|---|---|---|---|---|---|---|---|
| Sallam and Fürth [16] | | PINN | - | - | ✓ | - | ✓ |
| Maulik and San [17] | | Single layer NN | - | ✓ | - | ✓ | - |
| Zhou et al [27] | | - | ResNet | ✓ | - | ✓ | ✓ |
| Subramaniam et al [28] | Super resolution | - | GAN | ✓ | - | ✓ | - |
| Fukami et al [29] | | - | Multi Scale CNN | ✓ | - | ✓ | - |
| Kim et al [9] | | - | Cycle GAN | - | ✓ | ✓ | ✓ |
| Maulil et al [31] | | MLP | - | ✓ | - | ✓ | - |
| Pawar et al [32] | SGS inference | MLP | CNN | ✓ | - | ✓ | - |
| Gamahara and Hattori [33] | | Single layer NN | - | ✓ | - | ✓ | - |
| **Present work** | Super-resolution & SGS inference | - | Dual Scale ResNet | ✓ | - | ✓ | - |

In this paper, we aim to address the inability of conventional CNNs to reconstruct or infer the turbulent flow SubGrid Scales (SGS) with high-frequency content. Enhancing CNNs' capability to capture high-frequency turbulent features could enable high-fidelity turbulent CFD simulations with reduced computational cost, and potentially bridge the gap where Direct Numerical Simulation (DNS) becomes infeasible for real-world, high-Reynolds number engineering applications. We present a deep CNN Residual-Network (ResNet) with skip connections with dual scale convolution kernel to achieve this task. Table 1 shows a summary of the previous and present work contribution. Two CNN-based generative Residual Network (ResNet) models are implemented for Large Eddy Simulation Sub Grid Scale (SGS) resolving. The first model is a super-resolution model that can infer the high-resolution unresolved SGS velocity and vorticity fields from coarse grid LES inputs. The second model is a SGS stress tensor inference model which is trained to infer the SGS stress tensor from the resolved LES velocity and vorticity field inputs. The models are trained in supervised learning and prior analyses are implemented to test the models' predictions. The effect of using Single Scale Residual Block (SS-RB) and Dual Scale Residual Block (DS-RB) in the ResNet are studied on the learning and inference accuracy, and also the ability to infer the high-frequency features. The super-resolution model is trained and tested on LES 2-D planar injection jet flow problems at 2 different Reynolds numbers, two different jet configurations, and two upscale ratios. The SGS stress tensor inference model is tested on a classical DNS 2-D homogonous isotropic turbulence decay problem. The rest of the paper is structured as follows: In Section 2, the architectures of the Single/Dual Scale (SS/DS) Super Resolution Residual Network (SR-ResNet) and the Sub Grid Scale stress Residual Network (SGS-ResNet) are presented. In Sections 3 and 4, the super-resolution and the SGS stress tensor inference test cases are presented with the discussion. Last, in Sections 5 and 6, the conclusion and future work are discussed.

## 2. Methodology

In this section, the Residual Network (ResNet) is discussed, including the benefits of using Single Scale and Multi Scale Residual Blocks (SS-RB, MS-RB). In addition, two ResNet models are presented. The first is for Large Eddy Simulation (LES) super-resolution, and the second is for Sub Grid Scale Stress (SGS) tensor inference for LES.

### 2.1. Residual Networks (ResNet) and Residual Blocks (RB)

The Residual Network (ResNet) with skip connection was first introduced in 2015 by He et al [22] and made a significant breakthrough in the ML-based computer vision and image processing field; nowadays almost all CNN generative models for AI image/video are using their or extensions of their residual block and skip connection method. The main idea of the ResNet is forcing the CNN to learn the residual functions with reference to the input, instead of learning the unreferenced functions. This can be achieved by the skip connection between the Residual Blocks (RB) in any general NN. Learning the residual function with referenced input can overcome the network degradation problem [22],[26], and this mitigates or prevents the vanishing gradient phenomenon, the proof is presented in [34]. Hence, the Residual Blocks (RB) with skip connections allow designing much deeper networks. The broad overview idea of the ResNet with the skip connection is shown in Figure 1.

Figure 2 shows the Single Scale Residual Block (SS-RB) used in the ResNet architecture. Single Scale refers to that all convolution layers in the main path have a fixed convolution kernel size such as [3 × 3], regardless of the number of convolution layers appended in a single RB. The input to the $l$ layer RB is the output of the previous RB layer ($\mathbf{x}_l$). The RB operation applied on input ($\mathbf{x}_l$) is represented by $\mathcal{F}(\mathbf{x}_l, W_l)$, where $W_l$ are the trainable parameters to be optimized in training, and the output of the RB layer is ($\mathbf{x}_{l+1}$). Equation 9 shows the output of the RB layer where the last summation term $\mathbf{x}_l$ represents the identity mapping done by the skip connection. The skip connection path does not introduce any new trainable parameters, and the time required for the elementwise sum operation of the identity mapping is negligible. Multiple convolution layers can be stacked in series in the single scale RB; in Figure 2, two convolution layers exist. He *et al* [22] found that a single convolution layer in the RB does not add any advantage when compared to the plain CNN with no skip connections. Two or more convolution layers must be appended in series to make the RB effective.

The Batch Normalization (BN) is used to stabilize the training and prevent the model overfitting. De and Smith [35] showed that BN helps the ResNet to work at higher learning rates and achieve higher testing accuracy. On the other hand, authors in [36] found that the benefits of the BN are marginal for ResNet and only add training complexity. Hence, enabling the BN layer is problem-dependent. The nonlinear activation function RELU between the convolution layers helps the network to fit the nonlinear input features. In Figure 2, k, n, and s are the kernel size, number of output channels, and the convolution stride for the convolution operation, respectively.

The ResNet is modular and flexible, so it can be integrated into other architectures and can be used for broad applications. He *et al* [22] showed the success of the ResNet doing object detection on the MS COCO and PASCAL datasets, image segmentation on the MS COCO dataset, classification on ImageNet 2012 and CIFAR-10 datasets, and localization on the ImageNet 2012 dataset. For the Super Resolution Generator Adversarial Networks (SR-GAN) [24], the SR-ResNet is used as a generator.

Residual Blocks with Single Scale convolution kernels (SS-RB) performed well in general image super-resolution, classification, denoising, or deblurring tasks, Nevertheless, for broad spectrum feature image applications, the SS-RB struggled to reproduce the target images. For example, a generic human face has low-frequency features like nose, mouth, and eyes, and has high-frequency features like hair/skin texture. Multiscale convolution kernels in the Residual Blocks (MS-RB) were introduced to tackle this broad feature frequency challenge [37],[15]. In MS-RB, small kernel sizes ([1 × 1], [3 × 3]) are responsible for learning the high-frequency features, while the larger kernel sizes ([5 × 5], [7 × 7], [9 × 9]) are responsible for learning the low-frequency features. Turbulent flows have a broad spectrum feature as well, where large and small scales/features exist.

In the present work, we utilize a Dual Scale Residual Block (DS-RB) in the ResNet to perform super-resolution and infer the SGS stress tensor in LES. The DS-RB has two convolution layers in series for each convolution kernel scale path. The two convolution kernel sizes/scales are ([3 × 3], [5 × 5]). In addition, we employ the extra skip connections between different kernel scales inspired by Li *et at* [37], who applied the DS-RB for generic image super-resolution problems.

Figure 3 shows the DS-RB, where each scale path has the same layer order of the Single Scale Residual Block (SS-RB) shown in Figure 2, such as first convolution layer, followed by Batch Normalization (BN), RELU activation function, second convolution layer with double of feature map size, then BN. The bottom path is for the convolution kernel size [3 × 3] and the top one is for the convolution kernel size [5 × 5].

$a_1$ and $b_1$ are the extra skip connections implemented between the two convolution scale paths to share the extracted feature maps between [37]. The concentration block (Concat) combines the $a_1$ and $b_1$ tensors to the inputs of the second convolution layer for both scales. Similarly, the last Concat block combines the outputs of the dual scales convolution paths $a_2$ and $b_2$ into the last [1 × 1] bottleneck convolution layer. The [1 × 1] bottleneck convolution layer task is to reduce the feature map channels from 128 back to 64, so the identity mapping by elementwise summation can be implemented between the RB input ($\mathbf{x}_l$) and the RB operator output ($\mathcal{F}(\mathbf{x}_l, W_l)$), as shown in equation 9.

$$\mathbf{x}_{l+1} = \mathcal{F}(\mathbf{x}_l, \mathbf{W}_l) + \mathbf{x}_l \qquad (9)$$

## 2.2. Super Resolution ResNet (SR-ResNet) for turbulent flow

The first generative ResNet architecture used in the present work is a Super Resolution ResNet (SR-ResNet) [24] shown in Figure 4. The task of the SR-ResNet is to upscale the low-resolution LES filtered inputs $\mathbf{I}_{LR}$ to high-resolution outputs $\mathbf{I}_{HR}$ as shown in equation 10, where the $\mathcal{M}_{SR}$ is the super-resolution mapping operator of the SR-ResNet, and $\mathbf{W}$ is the set of trainable parameters to be optimized.

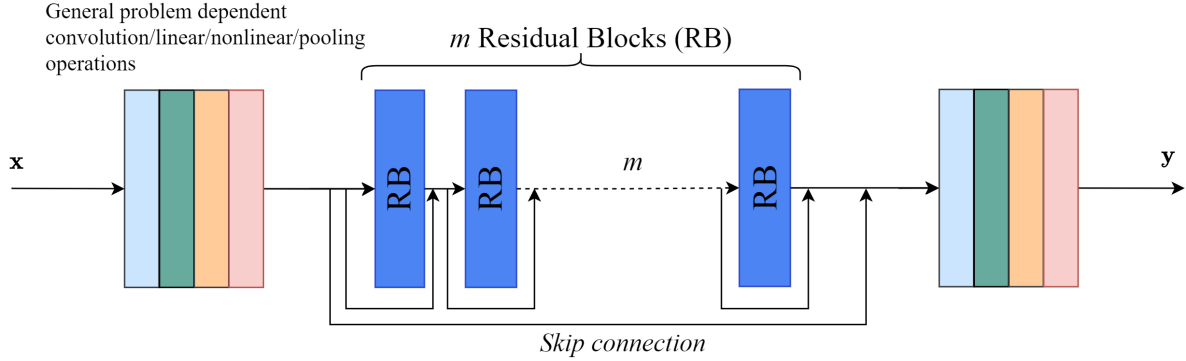

**Figure 1:** Basic Residual Network (ResNet) architecture for any general CNN problem. The input to the ResNet is passed through large kernel convolution layers, normalization layers, and nonlinear activation layers. Then the Residual Blocks (RB) are stacked to form a deeper CNN with skip connections in between to eliminate the vanishing gradient phenomena. A long skip connection is added along all RBs to maintain the information before the deep CNN.

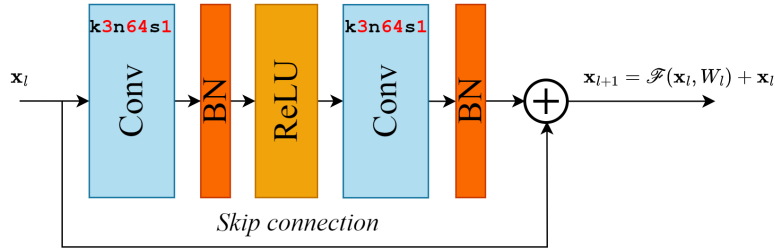

**Figure 2:** Single Scale Residual Block (SS-RB), both convolution layers are $3 \times 3$ kernel. The SS-RB has two Batch Normalization (BN) layers and a nonlinear ReLU activation function.

In Figure 4, the first $[9 \times 9]$ convolution layer creates a 64-channel feature map from the low-resolution input $\mathbf{I}_{LR}$. The large kernel size of this first convolution layer preserves the original information of the low-resolution input. A RELU activation function follows the $[9 \times 9]$ convolution to add a nonlinear learnable parameter mapping before the Residual Blocks (RB). $m$ RBs are stacked in series with skip connections in between, as discussed before. The RBs have lower convolution kernel size to reduce the computational cost. For the Single Scale RB (SS-RB), the convolution kernel size is $[3 \times 3]$, and for the Dual Scale RB (DS-RB) two convolution kernel sizes are $[3 \times 3]$ and $[5 \times 5]$.

The output of the RBs is passed through $[3 \times 3]$ convolution layer and Batch Normalization layer (BN) to stabilize/regulate the network, then by elementwise summation (ES) where it is added to the long skip connection tensor, as shown in Figure 4.

In SR-ResNet, the task of the path from the low-resolution input to the ES block is feature extraction through the convolution and nonlinear activation functions in the RBs. However, the resolution upscaling task is implemented in the Pixel Shuffle Blocks (PSB). The number of the PSB ($f$) determines the upscale factor, such that upscale factor $= 2^f$, or in other words, each PSB doubles the resolution of the input tensor. In the PSB, the pixelshuffler layer doubles the input resolution and decreases the number of channels by a factor of 4. Hence, the $[3 \times 3]$ convolution layer in the PSB has an output of 256 channels that goes down to 64 channels after a single pixelshuffler layer operation [38]. The last $[9 \times 9]$ convolution layer returns the feature map channels from 64 to the original problem dimension, 2-D or 3-D.

$$\mathbf{I}_{HR} = \mathcal{M}_{SR}(\mathbf{I}_{LR}, \mathbf{W}) \tag{10}$$

## 2.3. Sub Grid Scale ResNet (SGS-ResNet)

The second generative ResNet architecture used in the present work is a Sub Grid Scale (SGS) stress tensor inference (SGS-ResNet) as shown in Figure 5. The task of this network is to infer the SGS stress tensor $\tau_{ij}$ from

Omar Sallam et al.: *Preprint submitted to Elsevier* Page 7 of 33

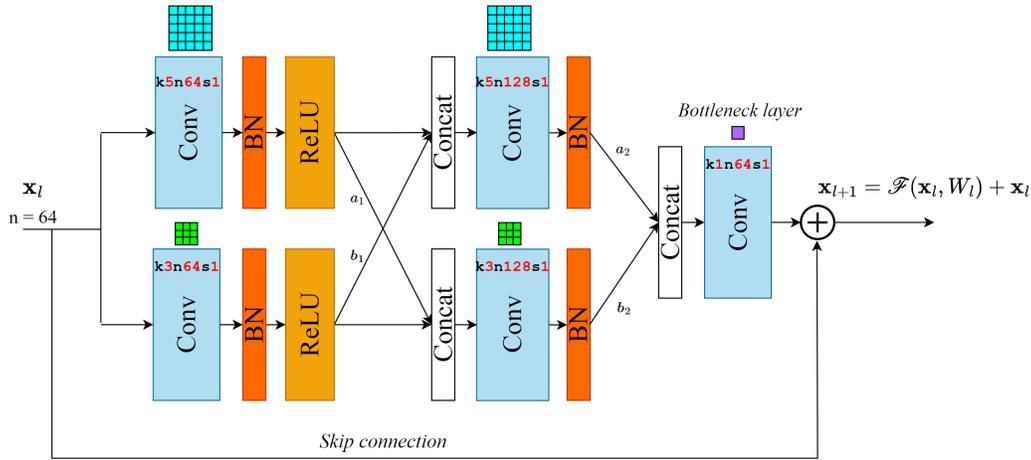

**Figure 3:** Dual Scale Residual Block (DS-RB), one path has convolution layers with $3 \times 3$ kernel, and the other path has convolution layers with $5 \times 5$ kernel. The information on the two paths is shared after each convolution layer.

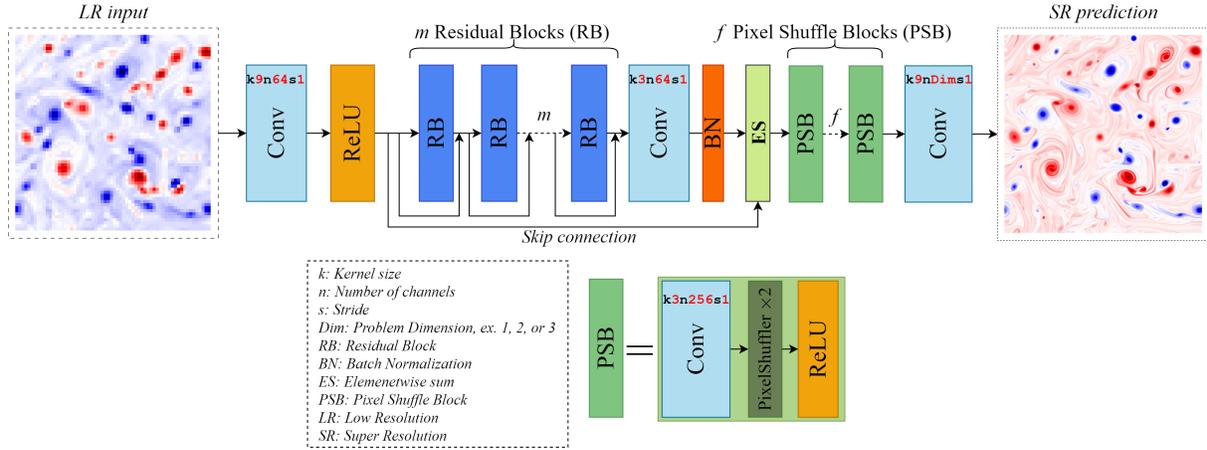

**Figure 4:** Super Resolution Residual Network (SR-ResNet) for Large Eddy Simulation. The network inputs are Low Resolution (LR) velocity fields and the output is a High Resolution (HR) one. The LR input is passed through a large kernel convolution layer [$9 \times 9$], followed by ReLU activation function, then $m$ Residual Blocks (RB). $f$ Pixel Shuffle (PS) layers are added to increase the layer output resolution. Each PS layer increase the resolution by factor of 2.

the low-resolution filtered velocity and vorticity fields $(u_{ij}, \omega_i)$ of LES as shown in equation 11. In equation 11, the $\mathbf{I}_{LR}$ includes the low-resolution filtered velocity and vorticity fields, $\mathbf{W}$ is the set of the trainable parameters to be optimized, and $\mathcal{M}_{SGS}$ is the mapping operator of the SGS-ResNet shown in Figure 5.

The architecture of the SGS-ResNet is similar to SR-ResNet, except there are no Pixel Shuffle Blocks (PSB) that upscale the input tensor resolution. The last [$9 \times 9$] convolution layer task is to reduce the feature map channels n number from 64 to 3 or 6 based on the problem dimension either 2-D or 3-D, respectively. For the present work, 2-D turbulence is implemented, hence the output channels n of the last convolution layer is 3 that represents ($\tau_{11}, \tau_{12}, \tau_{22}$).

$$\tau_{ij} = \mathcal{M}_{SGS}(\mathbf{I}_{LR}, \mathbf{W}) \tag{11}$$

Omar Sallam et al.: *Preprint submitted to Elsevier* Page 8 of 33

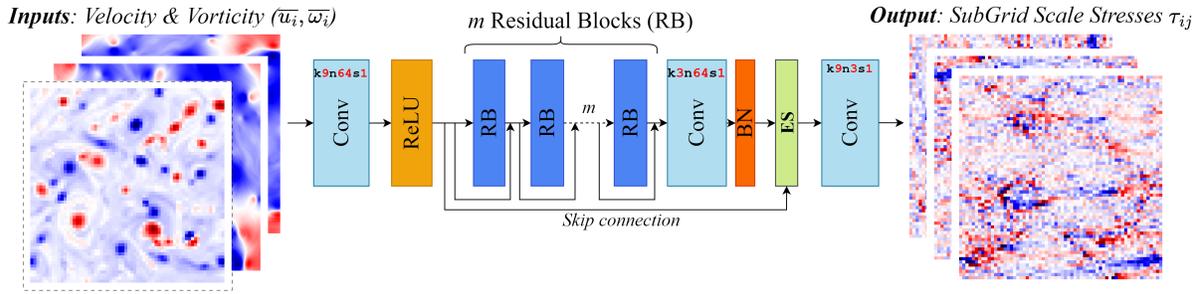

**Figure 5:** Sub Grid Scale (SGS) Residual Network (SGS-ResNet) for Large Eddy Simulation. The network inputs are velocity and vorticity fields, and the output is the SGS stress tensor $\tau_{ij}$. The network input and output tensor sizes are similar.

## 3. Super Resolution Implementation: A Planar Jet flow case study

In this section, both Super Resolution (SR) ResNet and DS-ResNet are tested for a two-dimensional planar jet injection LES problem. The SR network architecture is described in Figure 4, in the ResNet the Residual Block (RB) is a single scale as shown in Figure 2, while the RB of the DS-ResNet is a Dual Scale (DS) as shown in Figure 3.

Eight test cases are considered for this study as shown in Table 2, four test cases for the upscale ratio of 4 and four test cases for the upscale ratio of 8. The first two test cases for each upscale ratio are for single jet injection at Reynold's numbers $10^4$ and $2 \times 10^4$, while the last two test cases for each upscale ratio are for double jet injection at Reynold's numbers $10^4$ and $2 \times 10^4$.

Figure 6 illustrates the velocity profile for the single jet and double jet cases. The Reynold's number characteristic length is the jet nozzle height $H = 1$, $Re_H = \frac{U_\infty H}{\nu}$, $U_\infty = 1$ is the jet inlet velocity in $x_1$ direction. pimpleFoam [39], an incompressible flow solver in OpenFoam [40], is used to numerically simulate the LES problem. Smagorinsky model [5] is used to model the Sub Grid Scale (SGS) stresses. BlockMesh is used to generate the uniform mesh [300 × 100] square cells. Inlet velocity profiles for single and double jet injections are described in Figure 6, and walls have noSlip boundary conditions. The initial time step is 0.0005 seconds, and the maximum Courant number is 0.5.

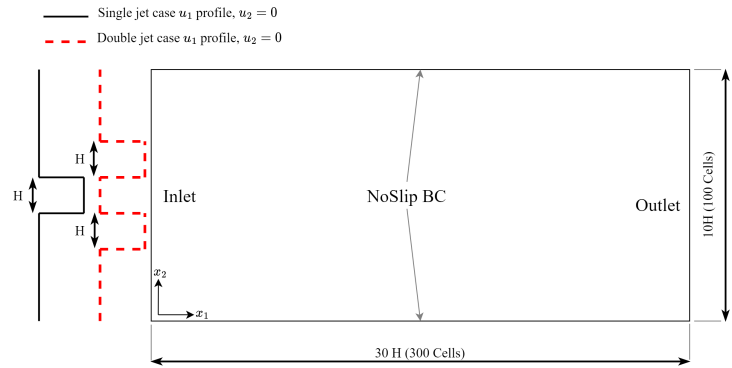

**Figure 6:** Planar jet injection geometry, mesh size, and boundary conditions. The solid black line is the velocity inlet profile for the single jet injection case, the dashed red line is the velocity profile for the double jet injection case.

### 3.1. Dataset preparation and Network parameters

All test cases in Table 2 are simulated for 1000 seconds and the velocity components are collected every 1 second giving a total input dataset of 1000 frames for the velocity components ($u_1, u_2$). Test case 1 and test case 5, single jet at $Re = 1 \times 10^4$, are used for training both the ResNet and the DS-ResNet while the other test cases are for testing. Since these networks are trained based on supervised learning, both Low Resolution (LR) inputs and the High Resolution (HR) outputs are paired for the training dataset. The output HR dataset, OpenFoam solution, is stored in a 4-D binary npy array to match the DataLoader utility in Pytorch, where the first index is the number of training frames =1000, the second index represents the number of channels =2 as only $(u_1, u_2)$ are the inputs, and the last two indices are 300,100 represent the input frame dimension. The LR dataset is attained by applying a sharp spectral filter to the HR dataset followed by a downsampling with factor 4 or 8 based on the target network SR upscaling factor. The input LR dataset is also a 4-D array with the same number of frames and channels but different in the last two indices of the frame dimensions based on the upscale ratio.

**Table 2**
Planar jet flow training and testing cases with the corresponding Reynold's number.

| Case number | Jet condition | Upscale ratio | Reynolds number ($Re_H$) |
|---|---|---|---|
| 1 | Single jet (**Train**) | 4 | $1 \times 10^4$ |
| 2 | Single jet (Test) | 4 | $2 \times 10^4$ |
| 3 | Double jet (Test) | 4 | $1 \times 10^4$ |
| 4 | Double jet (Test) | 4 | $2 \times 10^4$ |
| 5 | Single jet (**Train**) | 8 | $1 \times 10^4$ |
| 6 | Single jet (Test) | 8 | $2 \times 10^4$ |
| 7 | Double jet (Test) | 8 | $1 \times 10^4$ |
| 8 | Double jet (Test) | 8 | $2 \times 10^4$ |

For both ResNet and DS-ResNet, $m$ the number of Residual Blocks (RB) is set to 16, and $f$ the number of Pixel Shuffle (PS) blocks is 2 or 3 based on the desired upscaling factor 4 or 8 respectively. ADAMS algorithm [41] is used to minimize the network loss function and tune the network weights. The learning rate is 0.001, the batch size is 16, and the total number of iterations is $2 \times 10^4$.

The loss function to be minimized for both ResNet and DS-ResNet is a weighted average of the Mean Square Error (MSE) between the LES reference solution and the network predictions for the velocity field, vorticity field, and continuity equation residuals as shown in equation 12, where $\beta_{[\bullet]}$ is the weight for the corresponding loss metric.

$$L_{Generator} = \beta_u L_{(u_1,u_2)} + \beta_\omega L_{vorticity} + \beta_c L_{Continuity} \tag{12}$$

## 3.2. Super Resolution case study discussion

Super Resolution (SR) ResNet and DS-ResNet architectures are trained on cases 1 and 5 (single jet flow at $Re_H = 10^4$) for $2 \times 10^4$ iterations on an NVIDIA RTX A6000 GPU. The training is done for two upscaling ratios 4 and 8. For the 4 upscaling ratio, the number of Pixel Shuffle blocks $f = 2$, and for the 8 upscaling ratio, the number of Pixel Shuffle blocks $f = 3$.

Figures 7 and 8 show the training loss of velocity, vorticity, and continuity equation for the 4 and 8 upscaling ratios respectively. For both upscaling ratios, the DS-ResNet has faster decay compared to the ResNet for the three loss metrics. Figures 7 and 8 show that the impact of the DS-ResNet architecture on training accuracy improvement is more significant for the higher upscaling ratio 8.

Table 3 shows the number of trainable parameters, network size, training time, and the forward inference time for the SR-ResNet and SR-DS-ResNet architectures at two upscale ratios values 4 and 8. The Dual Scale Residual Block (DS-RB) increased the number of trainable parameters by 11 times, increased the network size by 15 times, increased the training time by 40 %, and increased the forward inference time by 185 %; the forward inference time is the time required to predict the HR velocity field from the LR inputs in milliseconds, the computed forward time is based on input size of [100 × 100] cells. It is observed that increasing the input LR grid size does not affect the forward time unless the GPU memory is about to reach its maximum allocation.

Figures 9 and 14 show the velocity and vorticity fields for one of the frames in the training and testing dataset respectively for upscale ratio 4, cases 1 and 3. The top row is the 4 times downsample filtered LES Low Resolution (LR), the second row for the reference LES solution, and the last two rows for the ResNet and DS-ResNet architectures predictions. Both architectures can reconstruct the HR field, however, the two figures show how the ResNet architecture is more blurry compared to the DS-ResNet architecture prediction that can reconstruct the high-frequency features in the velocity fields and the sharp shear layers interfaces. Figures 19 and 24 show the velocity and vorticity fields of the 8 times down sample LR input, LES, ResNet, and DS-ResNet predictions for cases 5 and 7. The DS-ResNet shows a better ability to reconstruct the sharp features of the velocity and vorticity fields compared to the ResNet. Figures 10 and 15 show a magnified vorticity fields of 4 times downsample case for one of the frames in the training and testing dataset respectively, while Figures 20 and 25 show a zoomed vorticity field of 8 times downsample case for one of the frames in the training and testing dataset respectively, cases 5 and 7. These figures show how coarse the vorticity field inputs are especially for the 8 times upscaling cases, but both the ResNet and the DS-ResNet can reconstruct the HR output. The DS-ResNet vorticity field predictions can better reconstruct the high-frequency and sharp features. Some of the magnified snapshots of the vorticity fields do not show clearly the improvement of the DS-ResNet effect compared to the ResNet, such as Figure 15, however by looking at statistical metrics such as mean error, error standard deviation, omnidirectional spectrum error, and bidirectional spectrum error, the improvement is obvious. These metrics are shown in Figures 16,17, 18, and will be discussed later in this section.

The velocity components prediction correlation, velocity error correlation, and the error correlation standard deviation between the reference LES and both 4 upscaling factor architectures ResNet and DS-ResNet are shown in Figures 11 and 16 for cases 1 and 3. The figures show that the slope of correlation of the DS-ResNet with the reference LES is higher than the slope of correlation of ResNet with LES for both training and testing datasets. In addition, the error slope correlation to the LES velocity and the ellipse of the error $3^{rd}$ standard deviation is lower for the DS-ResNet. Similarly, for upscale factor 8 the DS-ResNet has a higher correlation slope alignment with the LES velocity components, lower slope for error correlation, and lower error standard deviation, see Figure 21 and 26 for cases 5 and 7.

The omnidirectional energy spectrum and the omnidirectional energy spectrum relative absolute error for the ResNet and DS-ResNet predictions of 4 upscaling factor are shown in Figures 12 and 17 for the training and testing dataset respectively, cases 1 and 3. Both ResNet and DS-ResNet can reconstruct the omnidirectional energy spectrum back to the LES spectrum. On the right-hand side, the absolute relative omnidirectional error shows a higher error for the ResNet at higher wave numbers compared to the DS-ResNet architectures, this confirms that the Dual Scale Residual Block (DS-RB) is better at learning/reconstructing the high-frequency features. For more illustration, the 2-D spectrum error between the reconstructed velocity field and the LES velocity field for ResNet and DS-ResNet are shown in Figures 13 and 18 for the training and testing datasets, cases 1 and 3. The figures show how the 2-D spectrum error contours have lower values for the DS-ResNet architecture, especially at higher wave numbers. The same observation (lower 1-D and 2-D spectrum error for DS-ResNet architecture) is seen for the upscaling ratio 8 for both the training dataset and testing datasets, cases 5 and 7, as shown in Figures 22, 23, 27, and 28.

The summary of Mean Absolute Error (MAE) and the error standard deviation ($\sigma(E(U))$) are shown in Figure 29 and 30 for 4 and 8 upscaling factor test cases respectively. The figures show the effect of the DS-RB in accuracy and precision improvement for all training and testing cases at upscale ratios 4 and 8.

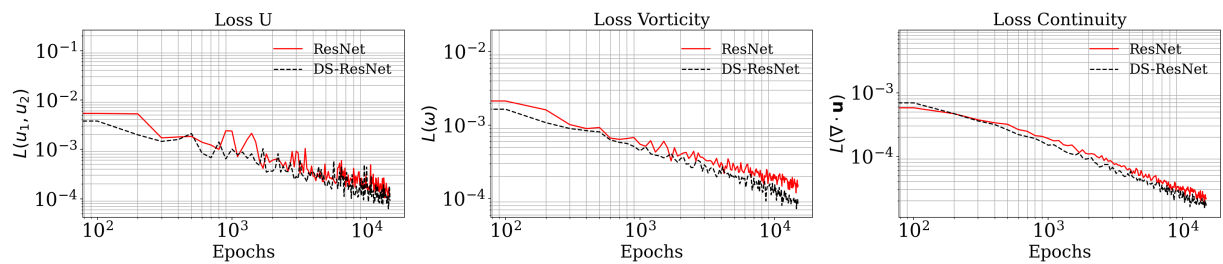

**Figure 7**: Generator training loss (Case1) for the ResNet and DS-ResNet architectures for the Super Resolution (SR) task by upscaling factor 4. The results show that DS-ResNet loss is decaying faster than the ResNet architecture.

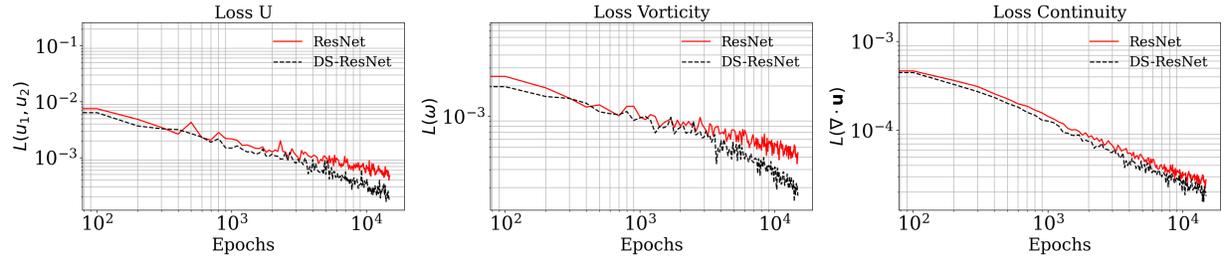

**Figure 8:** Generator training loss (Case 5) for the ResNet and DS-ResNet architectures for the Super Resolution (SR) task by upscaling factor 8. The results show that DS-ResNet loss is decaying faster than the ResNet architecture.

**Table 3**
Comparison between the Supper Resolution (SR) ResNet and DS-ResNet for number of trainable parameters, network size, training time, and the forward inference time at different numbers of Pixel Shuffle Blocks (PSB)

| Network | Upscale ratio | Trainable parameters | Network size [MB] | Training time [hr] | Forward time [ms] |
|---|---|---|---|---|---|
| SR-ResNet, $f = 2$ | 4 | $0.92 \times 10^6$ | 3.8 | 0.36 | 2.3 |
| SR-ResNet, $f = 3$ | 8 | $1.1 \times 10^6$ | 4.4 | 0.33 | 2.5 |
| SR-DS-ResNet, $f = 2$ | 4 | $11.7 \times 10^6$ | 47.3 | 0.5 | 6.7 |
| SR-DS-ResNe, $f = 3$ | 8 | $12 \times 10^6$ | 47.8 | 0.42 | 7.1 |

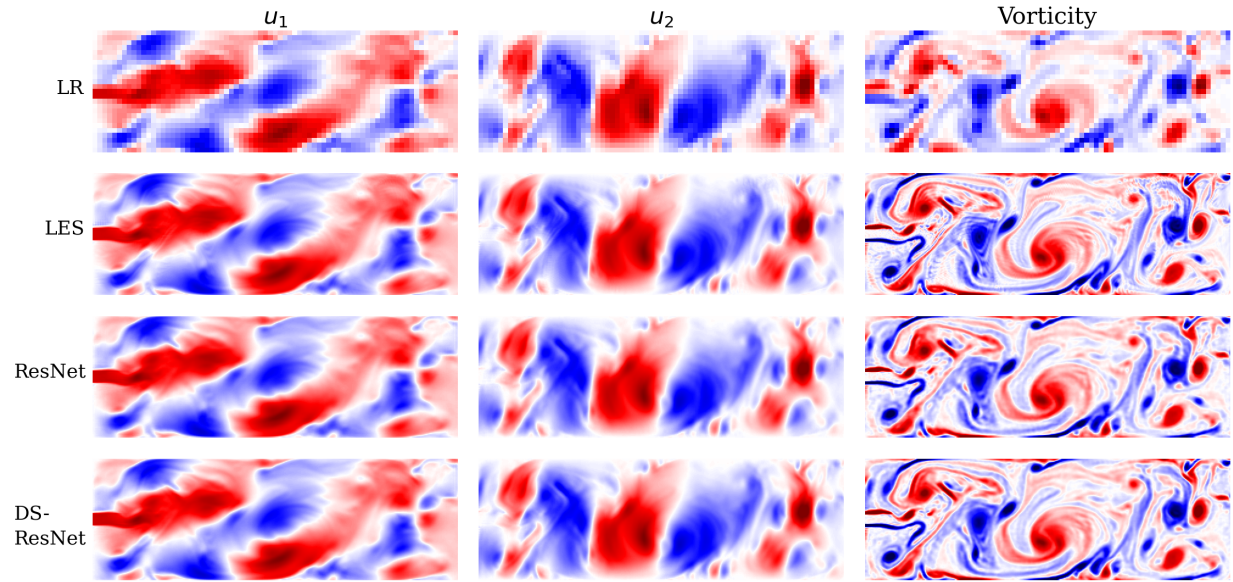

**Figure 9:** Case 1. Velocity and vorticity fields for Single jet at $Re = 1 \times 10^4$. The top row is the filtered LES with 4 downsampling ratio, the second row is the reference LES, third and fourth rows are for the ResNet and DS-ResNet predictions. See Figure 10 for magnified portions of the snapshot.

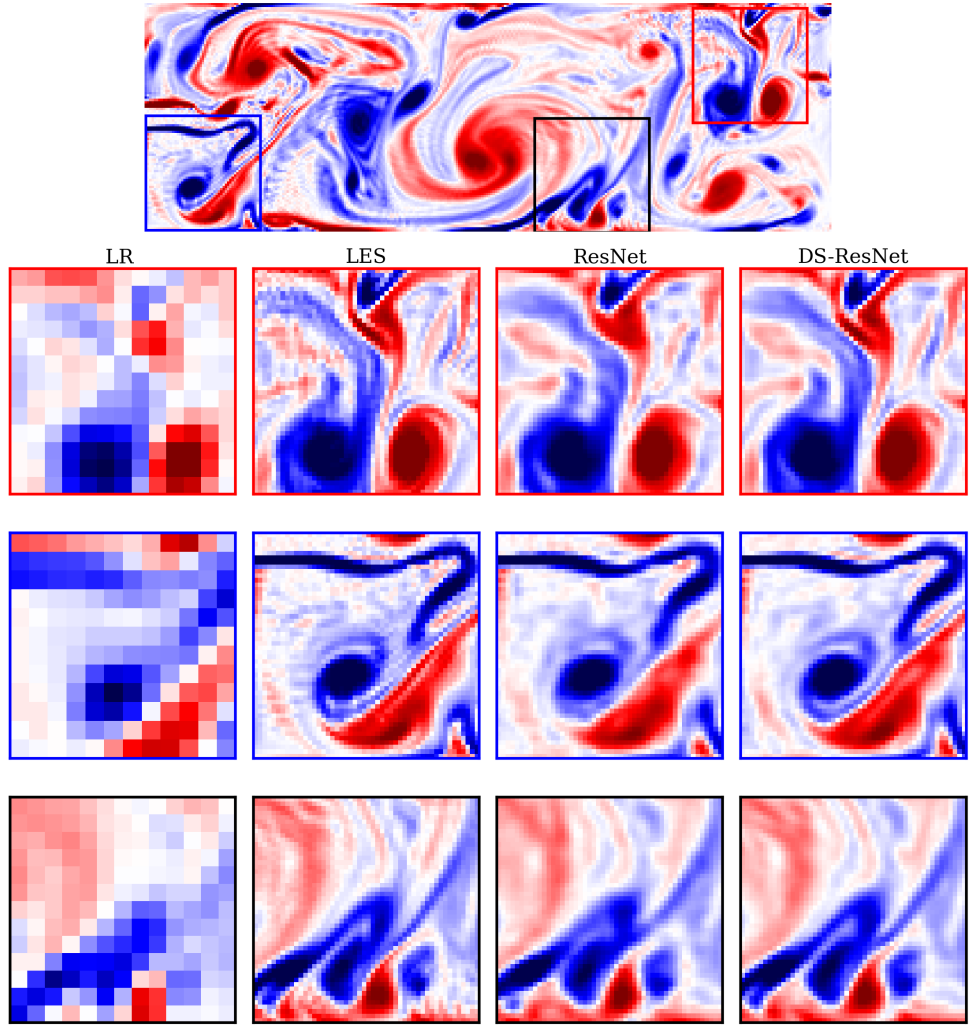

**Figure 10:** Case 1. Vorticity field for Single jet at $Re = 1 \times 10^4$. The left column is for the 4 ratio downsampling Low Resolution (LR), the second column is for the reference LES, third and fourth columns are for the ResNet and DS-ResNet predictions. DS-ResNet shows a better reconstruction for the high frequency features of the vorticity field compared to the ResNet architecture.

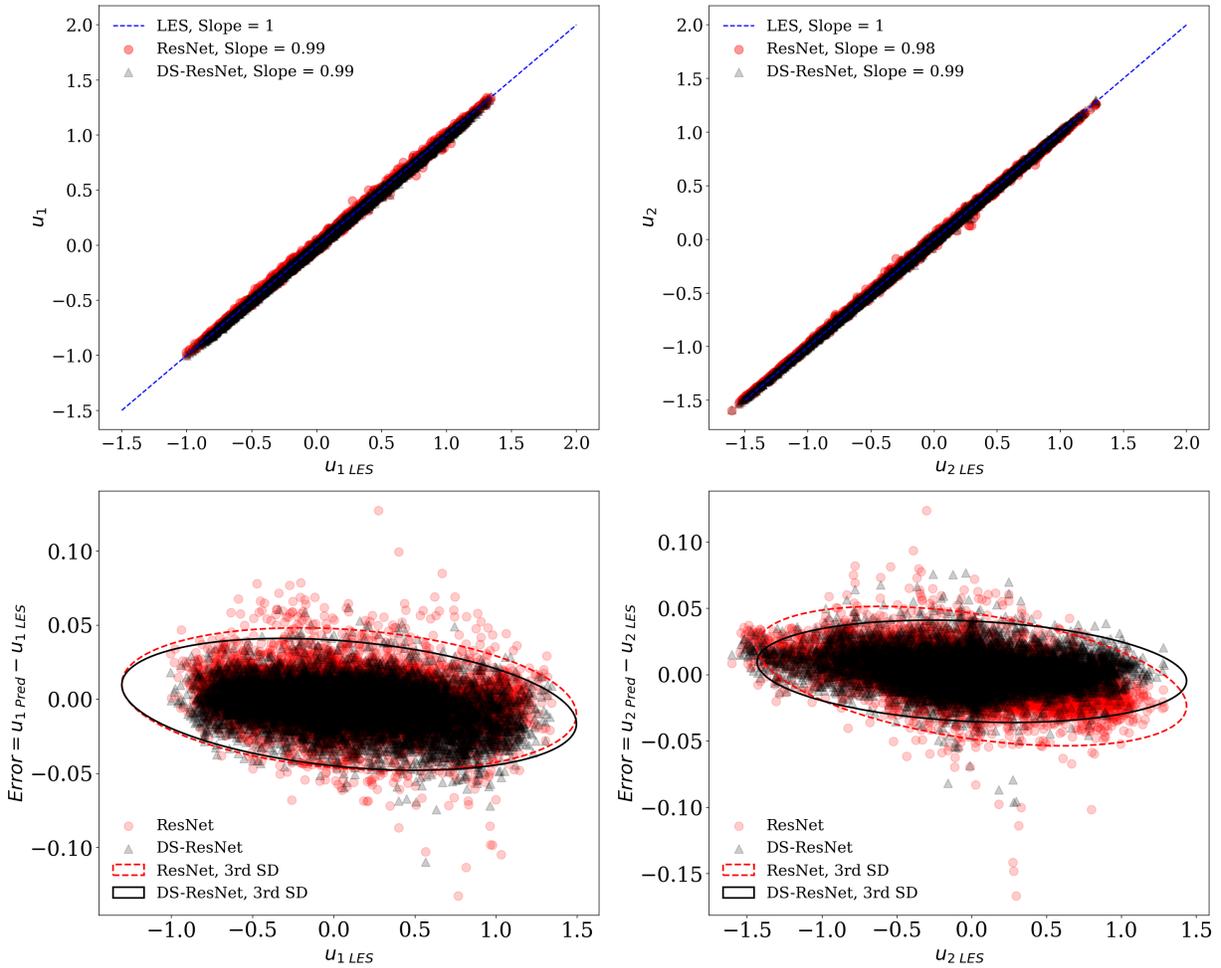

**Figure 11:** Case 1. Top: Velocity components correlation between the reference LES and both architectures ResNet and DS-ResNet predictions for 4 upscaling factor. Bottom: LES velocity component correlation with the ResNet and DS-ResNet prediction error and the error $3^{rd}$ standard deviation ellipse. The figure shows that DS-ResNet predictions have better alignment with LES, lower error slope, and lower error standard deviation compared to the ResNet architecture.

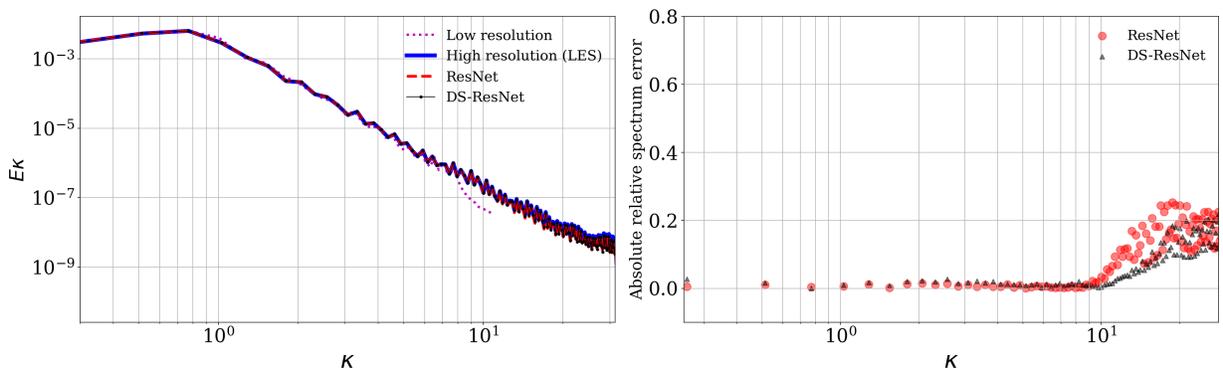

**Figure 12:** Case 1. Omnidirectional energy spectrum and absolute relative spectrum error for case 1 (Single jet at $Re = 1 \times 10^4$), with upscaling with factor 4. DS-ResNet has higher accuracy for the spectrum reconstruction at higher wave numbers compared to the ResNet

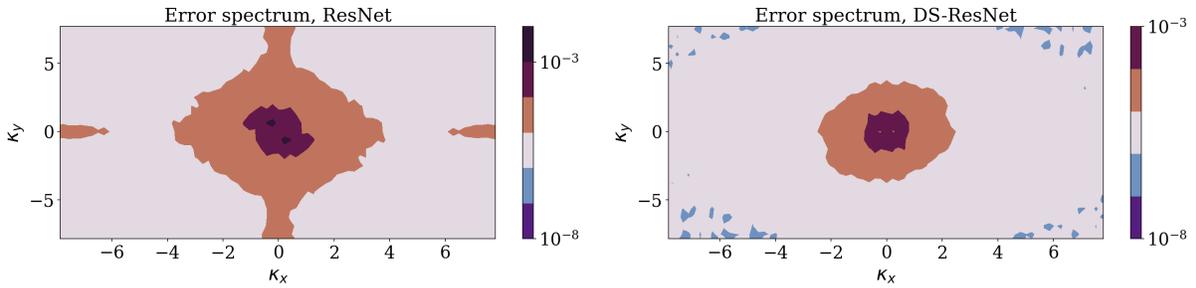

**Figure 13:** Case 1. 2-D energy spectrum error for case 1 (Single jet at $Re = 1 \times 10^4$), with upscaling with factor 4. The spectrum error is shrunk at higher wave numbers for the DS-ResNet architecture compared to the ResNet one.

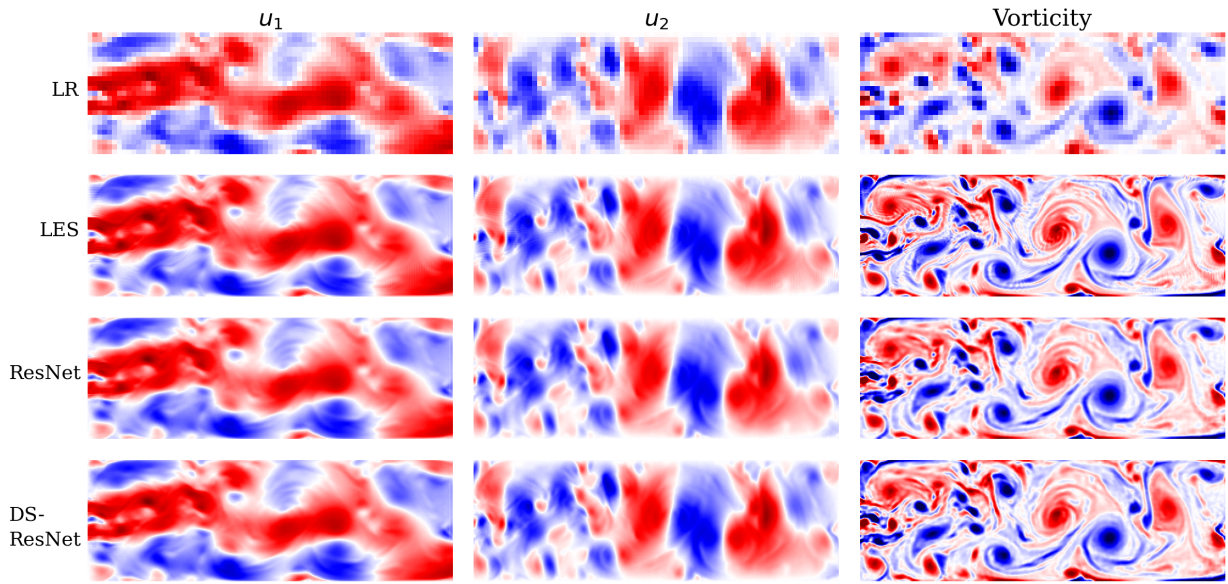

**Figure 14:** Case 3. Velocity and vorticity field Double jet at $Re = 1 \times 10^4$. The top row is the filtered LES with a 4 downsampling ratio, the second row is the reference LES, third and fourth rows are for the ResNet and DS-ResNet predictions. See Figure 15 for magnified portions of the snapshot.

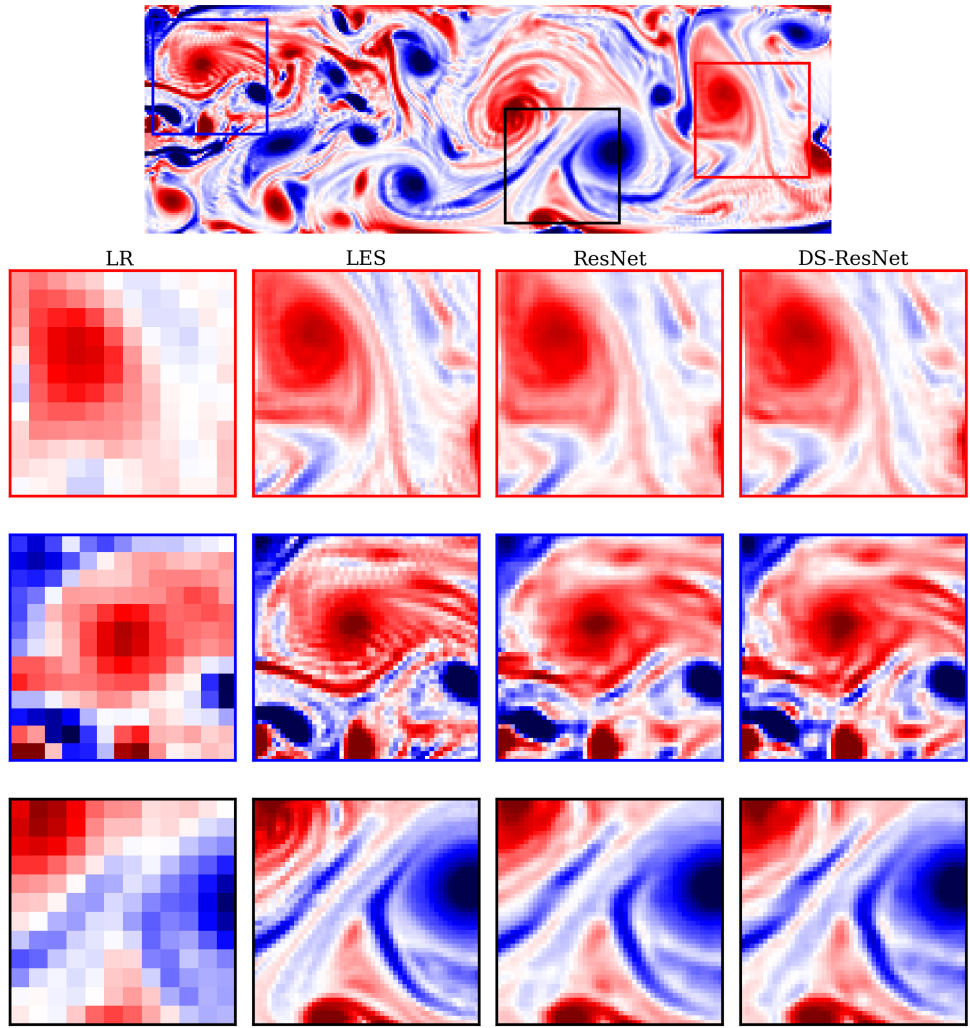

**Figure 15:** Case 3. Vorticity field for Double jet at $Re = 1 \times 10^4$. The left column is for the downsampling field by factor 4, the second column is the reference LES, third and fourth columns are for the ResNet and DS-ResNet predictions.

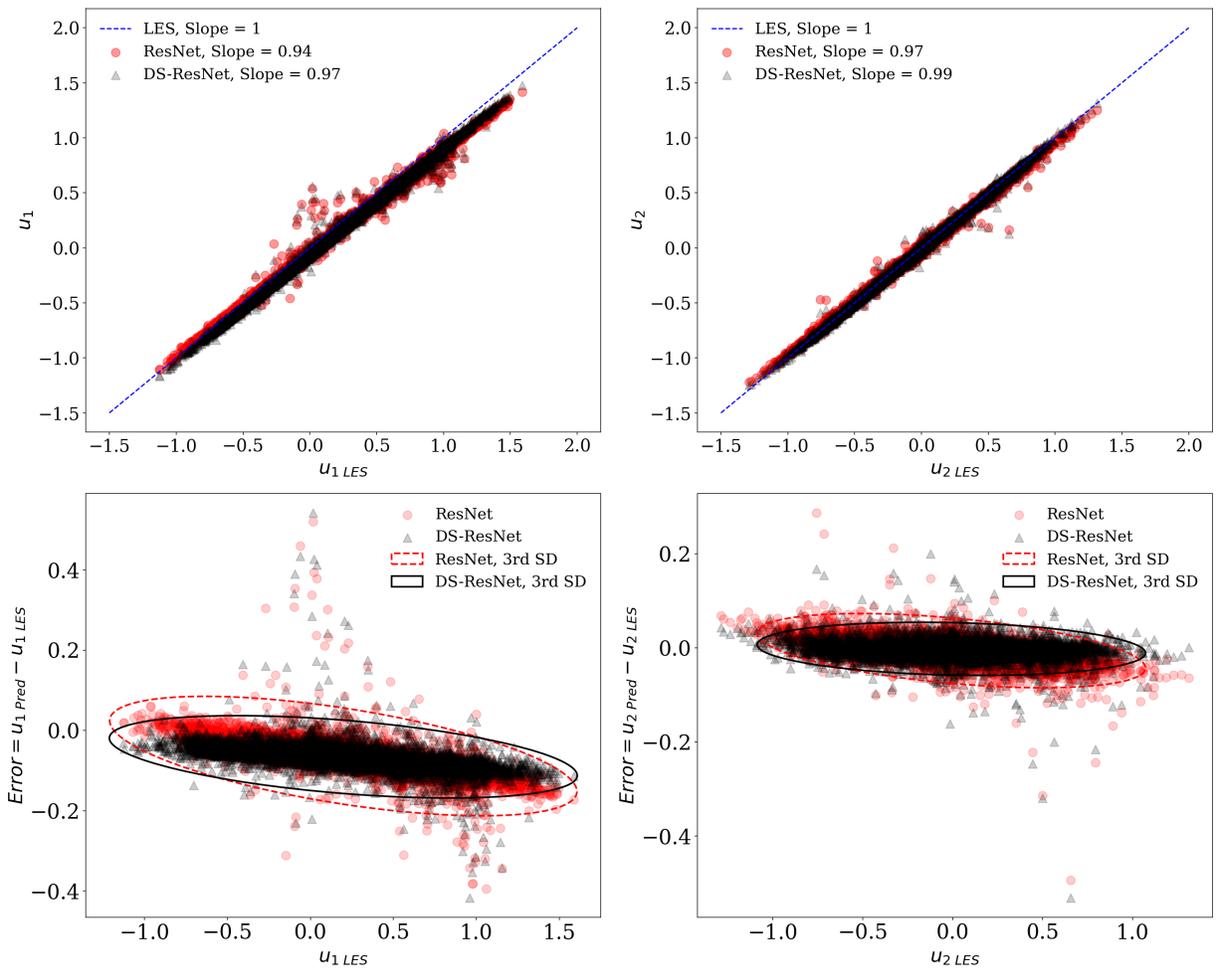

**Figure 16:** Case 3. Top: Velocity components correlation between the reference LES and both architectures ResNet and DS-ResNet predictions for 4 upscaling factor. Bottom: LES velocity component correlation with the ResNet and DS-ResNet prediction error and the error $3^{rd}$ standard deviation ellipse. The figure shows that DS-ResNet predictions have better alignment with LES, lower error slope, and lower error standard deviation compared to the ResNet architecture.

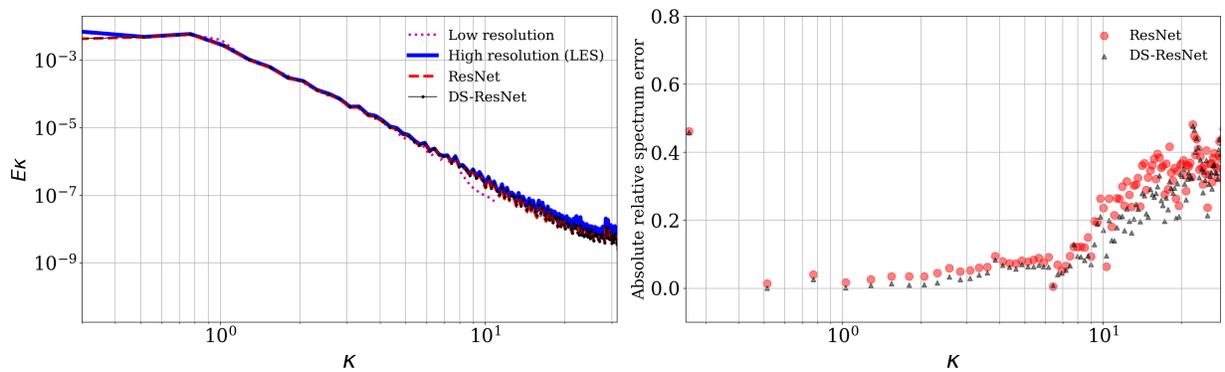

**Figure 17:** Case 3. Omnidirectional energy spectrum and absolute relative spectrum error for Double jet at $Re = 1 \times 10^4$. DS-ResNet has higher accuracy for the spectrum reconstruction at higher wave numbers compared to the ResNet

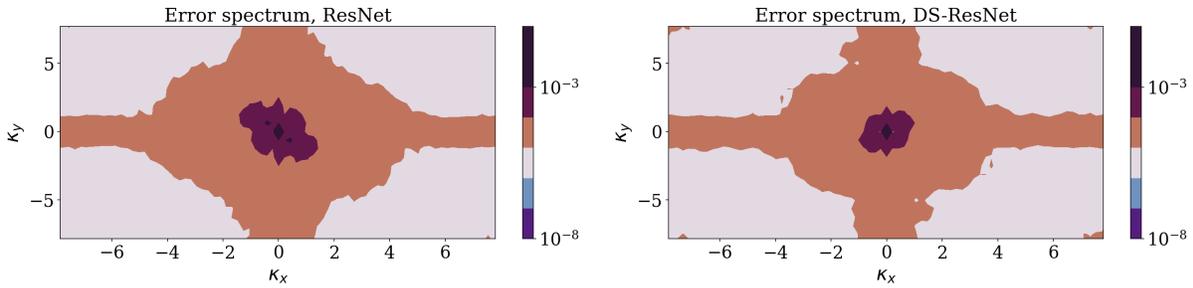

**Figure 18:** Case 3. 2-D energy spectrum error for Double jet at $Re = 1 \times 10^4$. The 2-D spectrum error is lower at high wave numbers for the DS-ResNet architecture compared to the ResNet one

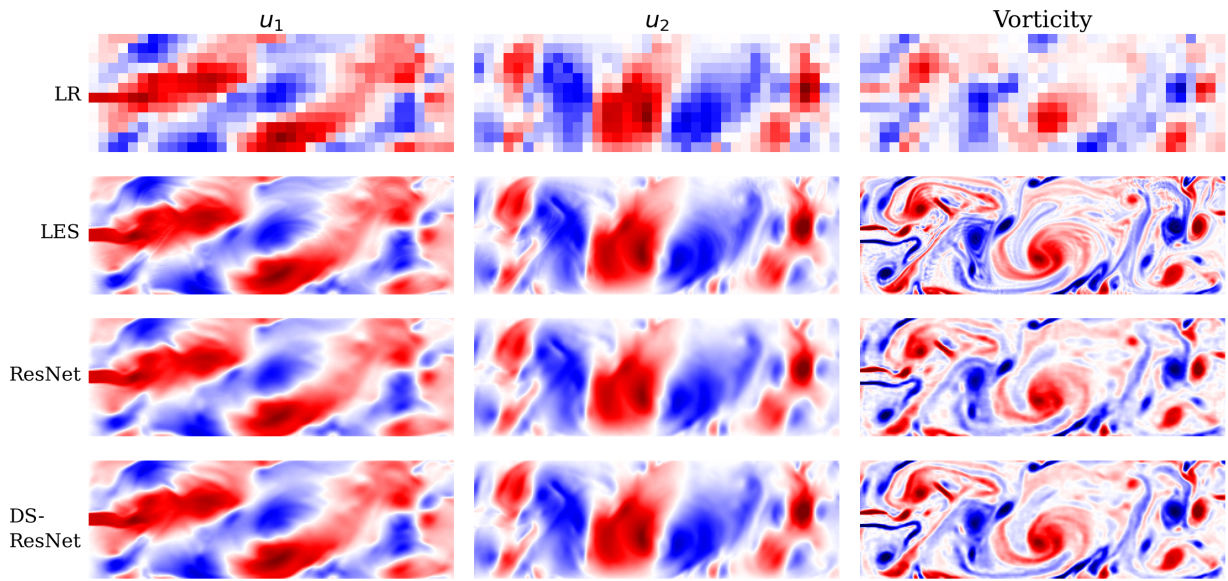

**Figure 19:** Case 5. Velocity and vorticity fields for Single jet at $Re = 1 \times 10^4$. The top row is the filtered LES with 8 downsampling ratio, the second row is the reference LES, third and fourth rows are for the ResNet and DS-ResNet predictions. See Figure 20 for magnified portions of the snapshot.

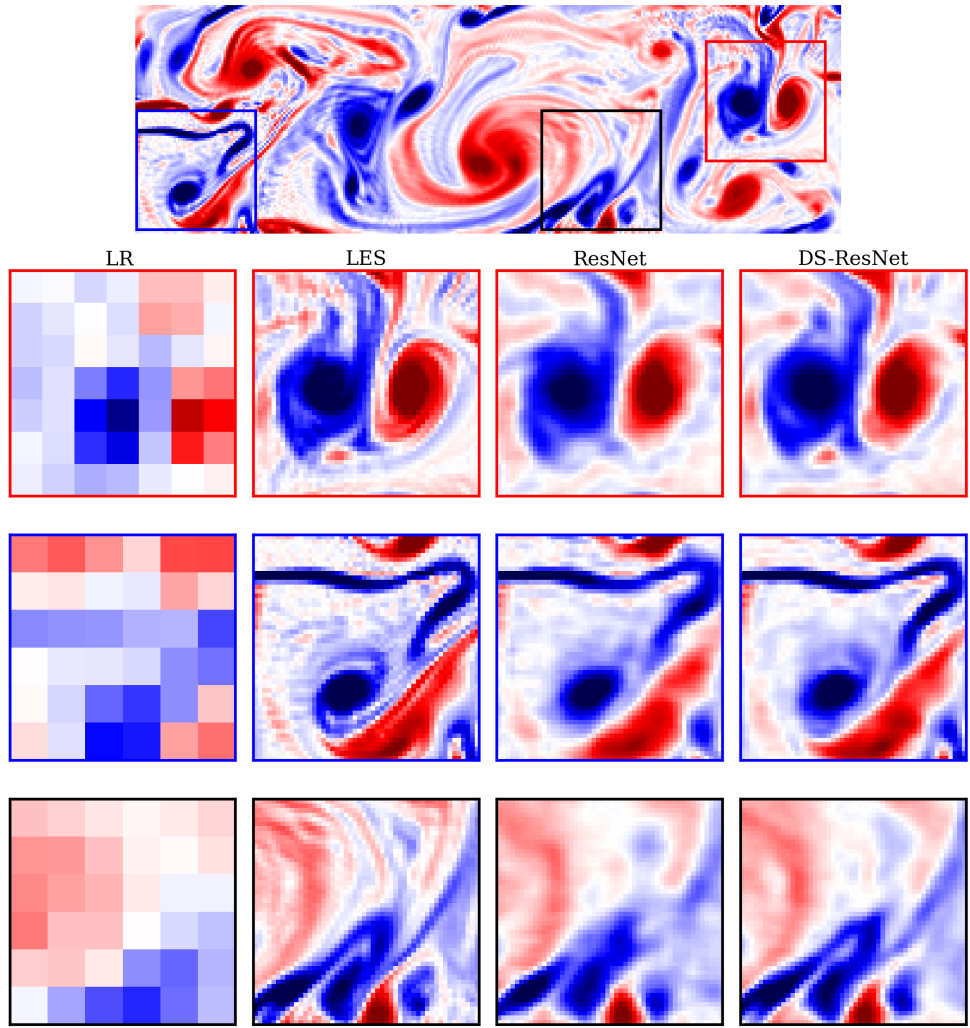

**Figure 20:** Case 5. Vorticity field for Double jet at $Re = 1 \times 10^4$. The left column is for the downsampling field by factor 8, the second column is the reference LES, third and fourth columns are for the ResNet and DS-ResNet predictions.

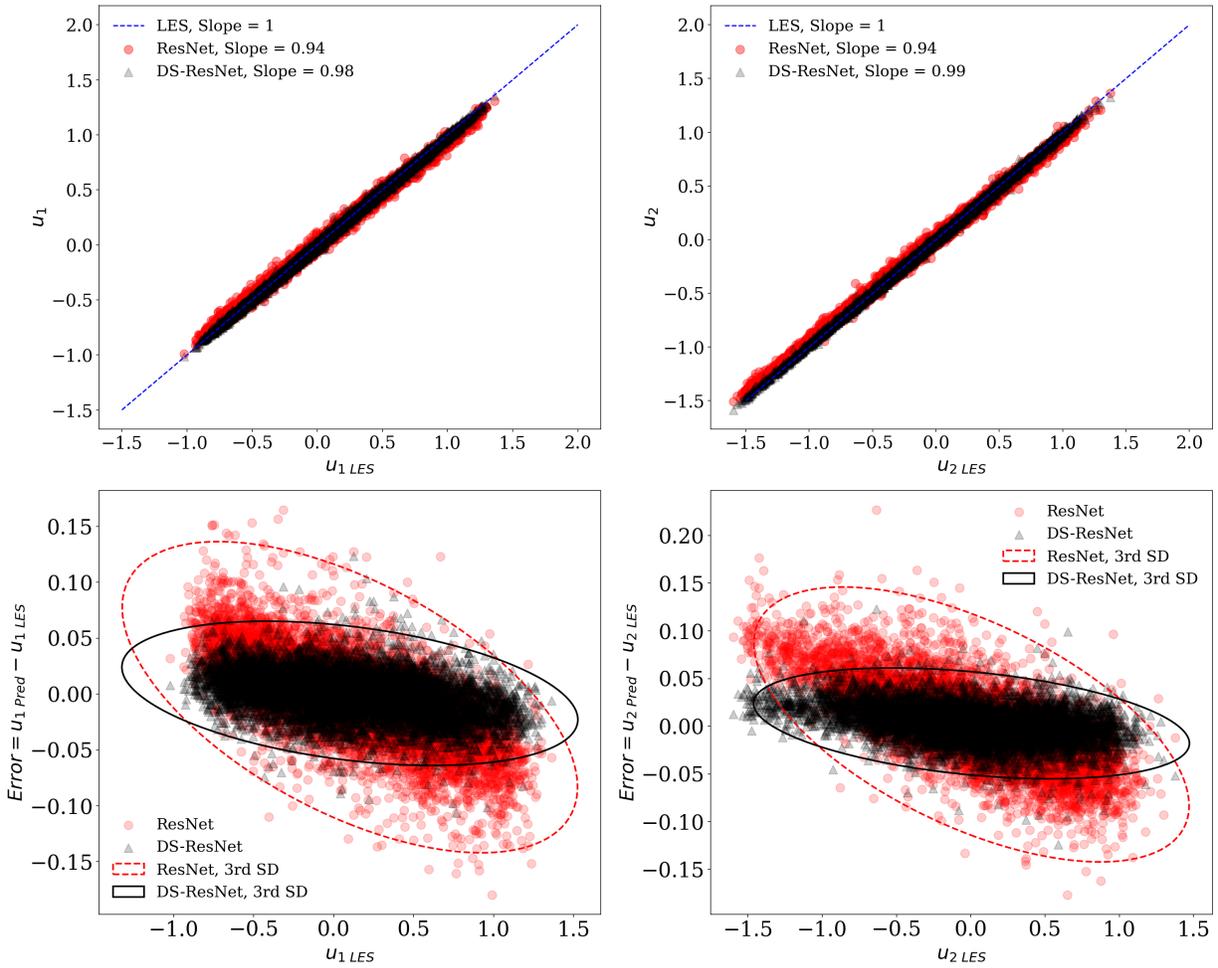

Figure 21: Case 5. Top: Velocity components correlation between the reference LES and both architectures ResNet and DS-ResNet predictions for 8 upscaling factor. Bottom: LES velocity component correlation with the ResNet and DS-ResNet prediction error and the error $3^{rd}$ standard deviation ellipse. The figure shows that DS-ResNet predictions have better alignment with LES, lower error slope, and lower error standard deviation compared to the ResNet architecture.

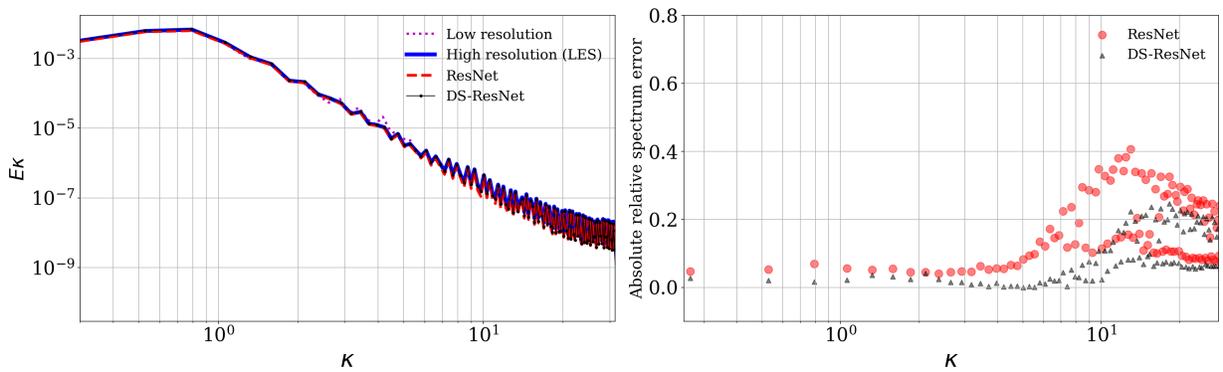

Figure 22: Case 5. Omnidirectional energy spectrum and absolute relative spectrum error for Single jet at $Re = 1 \times 10^4$ with upscaling with factor 8. DS-ResNet has higher accuracy for the spectrum reconstruction at higher wave numbers compared to the ResNet

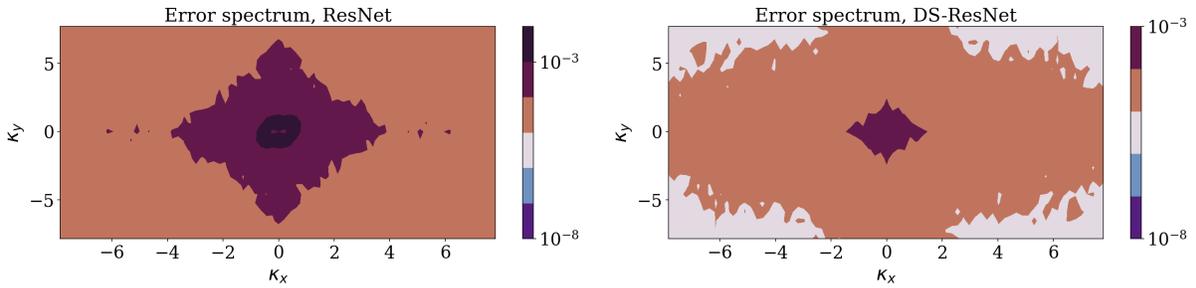

**Figure 23:** Case 5. 2-D energy spectrum error for Single jet at $Re = 1 \times 10^4$ with upscaling with factor 8. The spectrum error is lower at high wave numbers for the DS-ResNet architecture compared to the ResNet one.

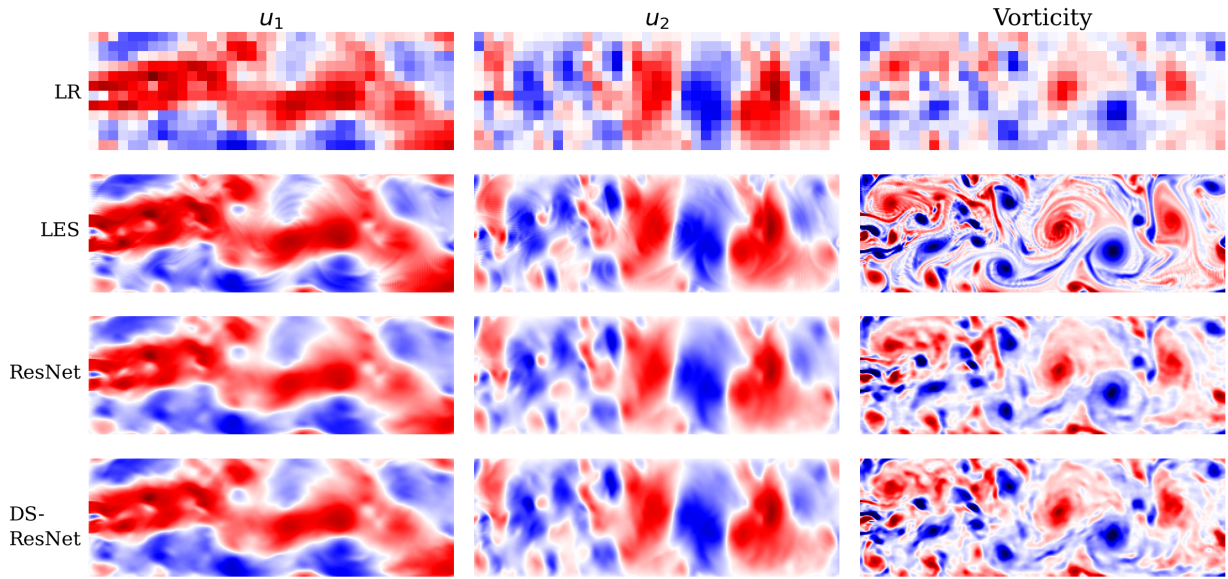

**Figure 24:** Case 7. Velocity and vorticity field for Double jet at $Re = 1 \times 10^4$. The top row is the filtered LES with a 4 downsampling ratio, the second row is the reference LES, third and fourth rows are for the ResNet and DS-ResNet predictions. See Figure 25 for magnified portions of the snapshot.

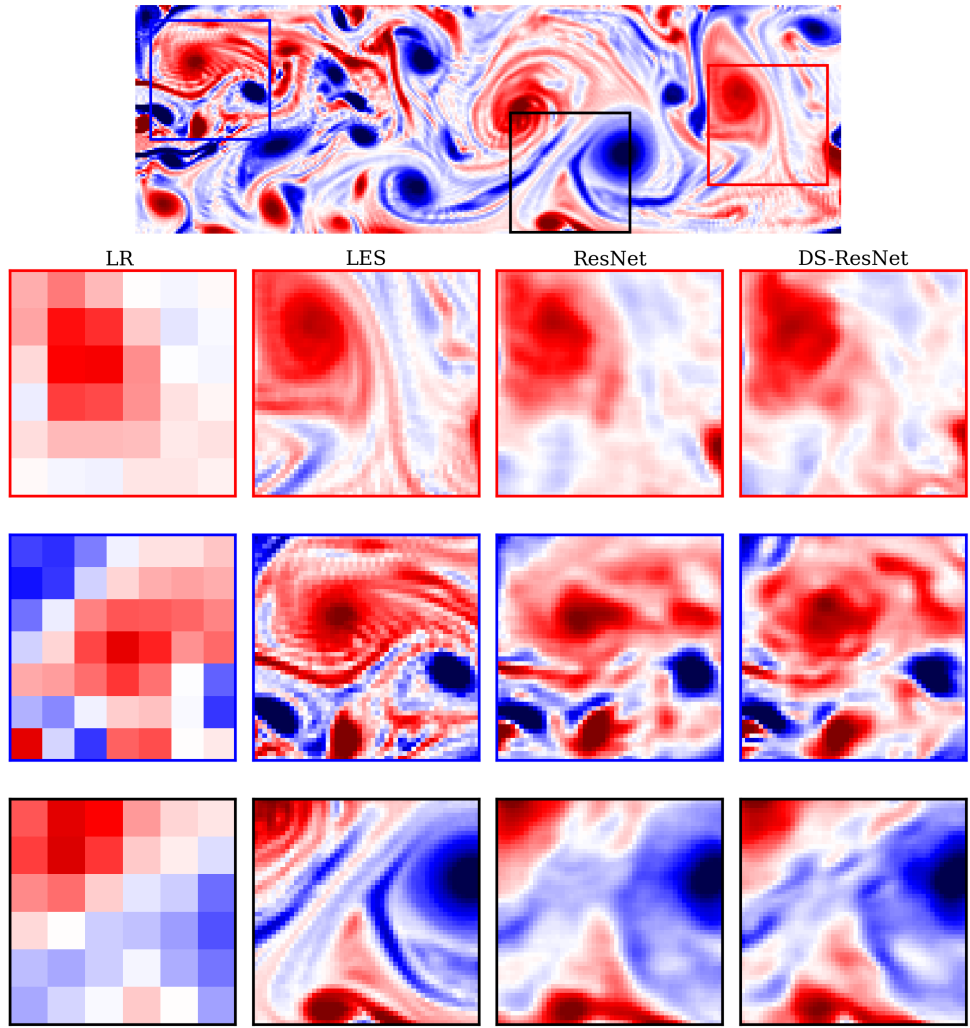

**Figure 25:** Case 7. Vorticity field for Double jet at $Re = 1 \times 10^4$. The left column is for the downsampling field by factor 8, the second column is the reference LES, third and fourth columns are for the ResNet and DS-ResNet predictions.

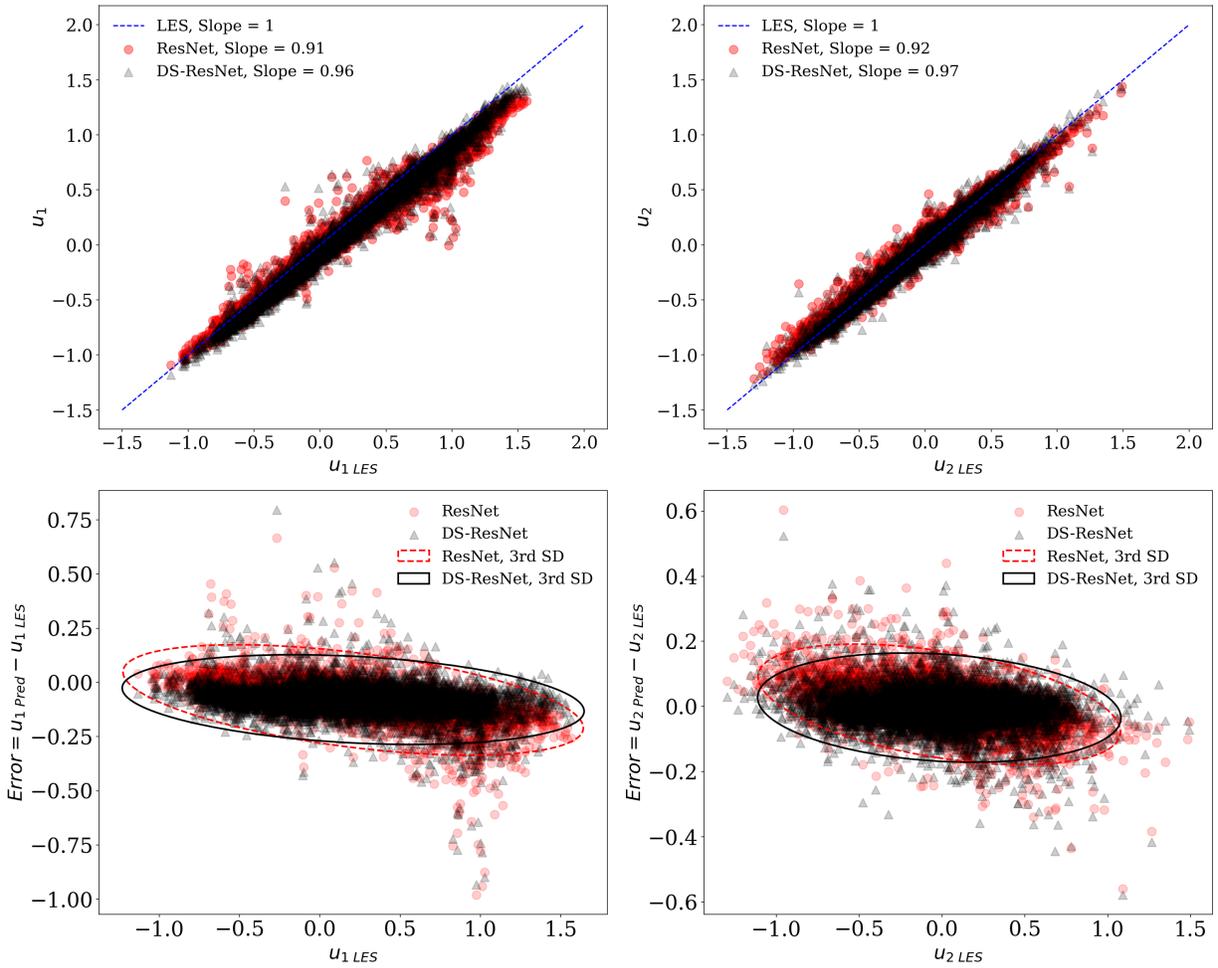

**Figure 26:** Case 7. Top: Velocity components correlation between the reference LES and both architectures ResNet and DS-ResNet predictions for 8 upscaling factor. Bottom: LES velocity component correlation with the ResNet and DS-ResNet prediction error and the error $3^{rd}$ standard deviation ellipse. The figure shows that DS-ResNet predictions have better alignment with LES, lower error slope, and lower error standard deviation compared to the ResNet architecture.

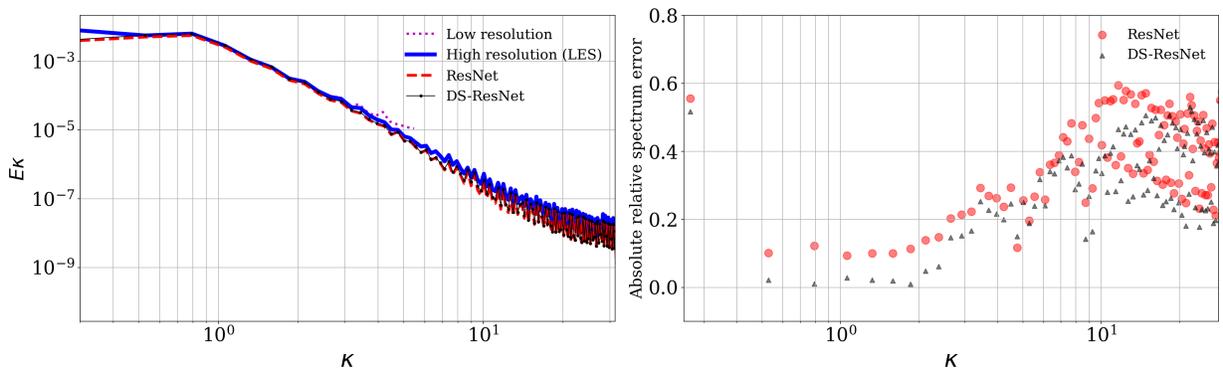

**Figure 27:** Case 7. Omnidirectional energy spectrum and absolute relative spectrum error for Double jet at $Re = 1 \times 10^4$ with upscaling with factor 8. DS-ResNet has higher accuracy for the spectrum reconstruction at higher wave numbers compared to the ResNet

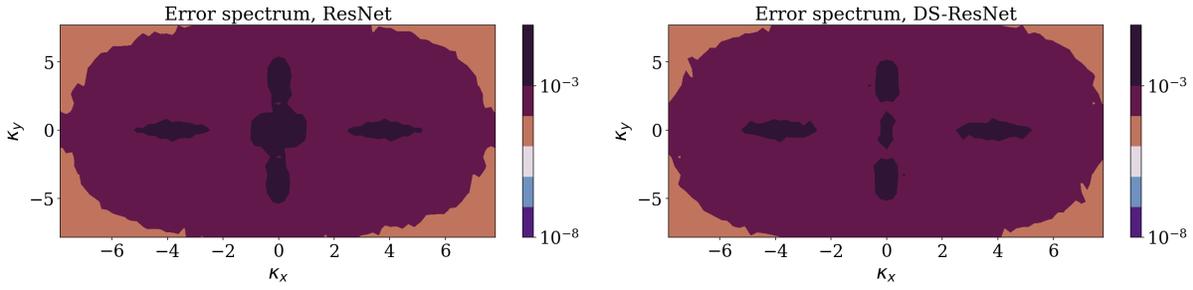

**Figure 28:** Case 7. 2-D energy spectrum error for Double jet at $Re = 1 \times 10^4$ with upscaling with factor 8. The spectrum error is lower at high wave numbers for the DS-ResNet architecture compared to the ResNet one.

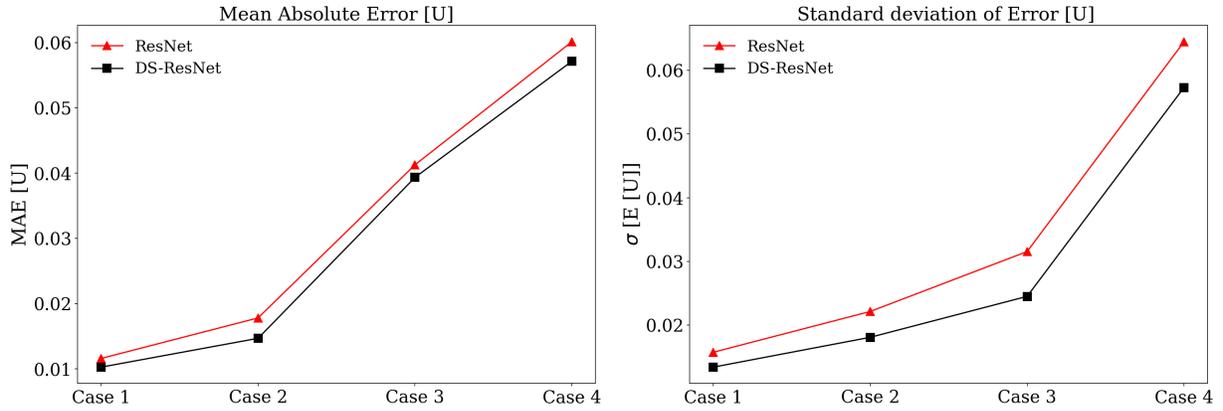

**Figure 29:** Mean Absolute Error (MAE) and the error standard deviation between the HR reference LES solution and the ResNet and DS-ResNet predictions with factor 4 downsample LR inputs for all cases, see Table 2. DS-ResNet has lower error and lower error standard deviation than the ResNet architecture.

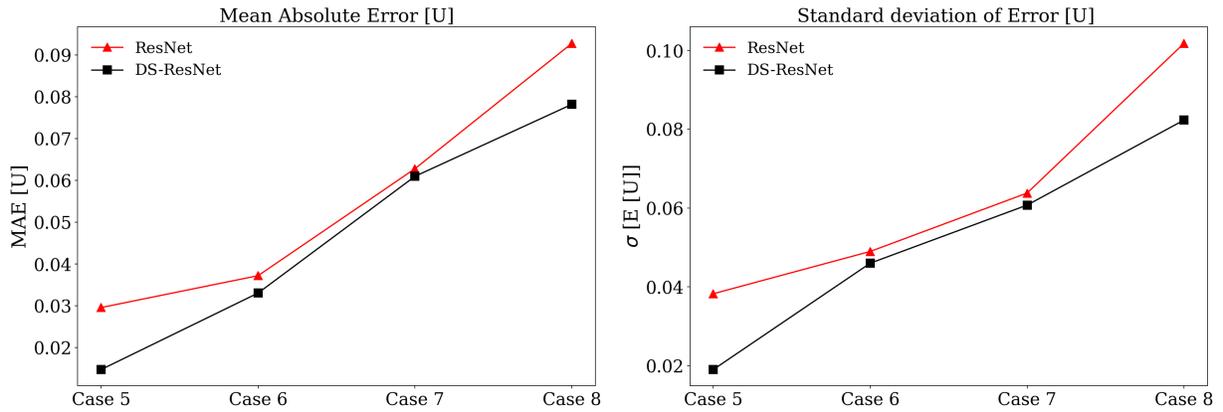

**Figure 30:** Mean Absolute Error (MAE) and the error standard deviation between the HR reference LES solution and the ResNet and DS-ResNet predictions with factor 8 downsample LR inputs for all cases, see Table 2. DS-ResNet has lower error and lower error standard deviation than the ResNet architecture.

## 4. Sub Grid Scale (SGS) stress inference: 2-D isotropic homogeneous turbulence decay case study

In this section, both ResNet and DS-ResNet are tested for Sub Grid Scale (SGS) stress tensor inference task of a 2-D homogeneous isotropic decay turbulence that follows the Kraichnan-Batchelor-Leith (KBL) theory [42, 43]. The generative SGS stress inference architecture is described in Figure 5, for ResNet, the Residual Block (RB) is a single scale as shown in Figure 2, while in DS-ResNet, the RB is a Dual Scale (DS) as shown in Figure 3. A Direct Numerical Simulation (DNS) is implemented to simulate the incompressible decay turbulence problem using pimpleFoam numerically [39], an OpenFoam [40] incompressible flow solver. BlockMesh utility is used to generate the square geometry with a dimension of $[2\pi \times 2\pi]$ and uniform square grid $[1024 \times 1024]$ cells. The time step is $10^{-4}$ and the maximum Courant number is 0.7. The mesh size and the time step are selected following previous studies simulating the same problem with the same initial Reynolds number range [44, 45, 46, 47].

All walls have periodic boundary conditions. The initial velocity field is computed by the Inverse Fast Fourier Transform (IFFT) from the energy spectrum shown in Equation 13 following [48]. In equation 13, $\kappa$ is the wave number magnitude. $\kappa_p$ is the dominant wavenumber where the initial energy spectrum has its peak value. For the present work $\kappa_p = 12$, $E_a$ is a normalization constant to control the initial root mean square velocity magnitude $u_{rms}$, for more details, see [49]. The case is simulated at initial Reynold's number $Re_0 = 25000$, where $Re_0 = \frac{u_{0,rms}\lambda_p}{\nu}$ is based on the dominant wavenumber $\lambda_p = \frac{2\pi}{\kappa_p}$ and the initial root mean squared velocity field $u_{0,rms}$.

The problem is simulated for 50 seconds; velocity and vorticity data are recorded every 0.2 seconds resulting in a dataset with 250 frames. The evolution of the vorticity contours is shown in Figure 31. The energy spectrum $E(\kappa)$ of the DNS is proportional to $\kappa^{-4}$ in the inertial range that agrees with the local theory of 2-D turbulence [50, 44]. The supervised training dataset is 70 % of the available frames [51], hence the total number of training and testing frames are 175 and 75 respectively.

$$E(\kappa) = E_a \left(\frac{\kappa}{\kappa_p}\right)^4 e^{-2\left(\frac{\kappa}{\kappa_p}\right)^2} \tag{13}$$

### 4.1. Dataset preparation and Network parameters

The input dataset for the SGS stress prediction networks is the Filtered DNS (FDNS) velocity components and the vorticity fields. The FDNS fields are obtained by applying a sharp spectral filter to the DNS data and then downsampling. For this test case, the DNS resolution is $[1024 \times 1024]$ while the FDNS resolution is $[64 \times 64]$, see Figure 32. The input dataset becomes a 4-D npy binary array. The first dimension is the number of training frames; The second dimension is 3 which represents the number of input channels $(u_1, u_2, \omega)$; the last two dimensions are 64 which represents the input frame resolution.

The output training dataset is the true DNS SGS stresses $\tau_{ij,DNS}$ computed from equation 5; the output training dataset is a 4-D npy binary array. The first dimension is the number of training dataset frames, the second dimension is 3 that represents $(\tau_{11}, \tau_{12}, \tau_{22})$, and the last two dimensions are the 64 representing the resolution of the output frames. For both ResNet and DS-ResNet, $m$ the number of Residual Blocks (RB) is set to 16. ADAMS algorithm [41] is used to minimize the network loss function and tune the network weights. The learning rate is 0.001, the batch size is 25, and the total number of iterations is $5 \times 10^4$. The loss function to be minimized for both ResNet and DS-ResNet is the Mean Square Error (MSE) between the DNS SGS Stresses $\tau_{ij,DNS}$ and the Networks SGS stresses $\tau_{ij,Pred}$ predictions

### 4.2. SGS stress prediction case study discussion

SGS stress tensor inference ResNet and DS-ResNet architectures are trained on the same training dataset for $5 \times 10^4$ iterations on an NVIDIA RTX A6000 GPU. DS-ResNet showed a dramatically faster learning trend compared to the ResNet. Figure 33 shows the generator loss $L_G$ for both architectures. Table 4 shows that Dual Scale Residual Block increases the number of the trainable parameters and the network size by about 15 times, while the increase of the training time is only 40%. The last column of the table shows the forward time in milliseconds, the time required to make SGS stress prediction from velocity and vorticity fields inputs. The forward time for the DS-ResNet architecture is 3 times the time for the ResNet model.

Figure 34 shows the SGS stress tensor for one of the frames in the testing dataset. The top row is the true DNS SGS stresses, the second row for the Smagorinsky model evaluated from equation 6, and the last two rows for the

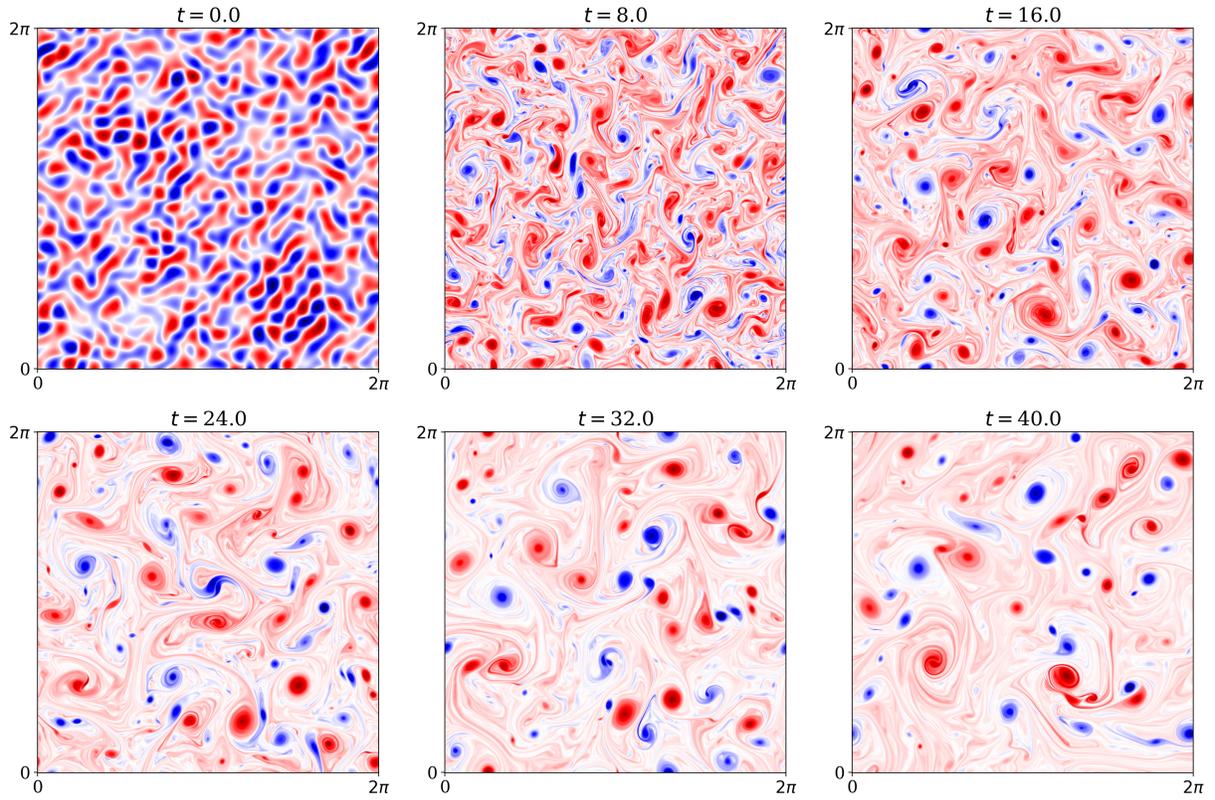

**Figure 31:** Vorticity evolution contours of DNS the incompressible isotropic homogenous free decay turbulence at initial Reynolds number $Re_0 = 25000$ and grid size $= [1024 \times 1024]$

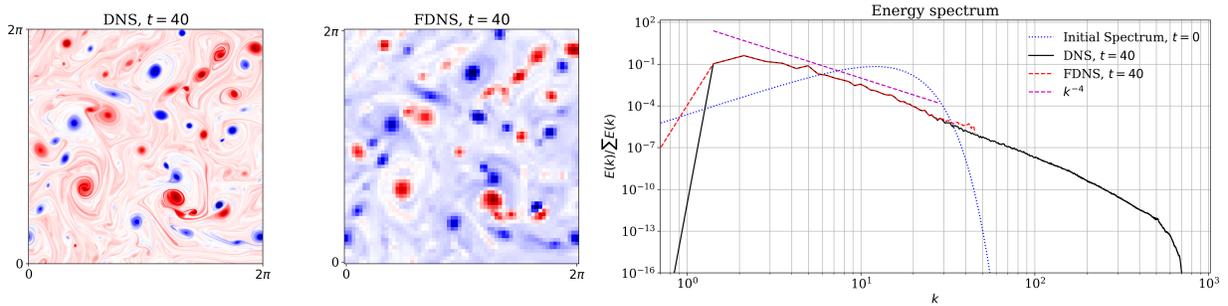

**Figure 32:** DNS and Filtered DNS (FDNS) vorticity contours and omnidirectional energy spectrum for the incompressible isotropic homogonous free decay turbulence.

ResNet and DS-ResNet predictions respectively. Smagorinsky SGS stress is very far from the true DNS SGS stresses, while both ResNet and DS-ResNet qualitatively align with the DNS SGS stresses. The figure shows that DS-ResNet can better reconstruct SGS stresses compared to ResNet which has artifact effects; this is more obvious for the $\tau_{11}$, the visual qualitative improvement for $\tau_{22}$ is not clearly seen in the Figure, however, the statistical comparison, discussed later, shows how the DS-ResNet improved the prediction. Figure 35 shows the Probability Density Function (PDF) of the SGS stresses for the true DSN, Smagorinsky, ResNet, and DS-ResNet. The PDF for both ResNet and DS-ResNet align with DNS $\tau_{ij}$, while the Smagorinsky PDF is more concentrated at the center.

**Table 4**
Comparison between the Sub Grid Scale (SGS) stress prediction ResNet and DS-ResNet architectures for a number of trainable parameters, network size, training time, and the forward inference time.

| Network | Number of trainable parameters | Network size [MB] | Training time [hr] | Forward time [ms] |
|---|---|---|---|---|
| SGS-ResNet | $0.7 \times 10^6$ | 3 | 0.49 | 2.1 |
| SGS-DS-ResNet | $11.5 \times 10^6$ | 46.1 | 0.68 | 6.6 |

The joint distribution between the SGS stress predictions and the true DNS SGS stresses is shown in Figure 36. The joint PDF shows that DS-ResNet SGS stress tensor predictions are more aligned with true DNS SGS stress tensors compared to the ResNet and the Smagorinsky. The Mean Absolute Error (MAE) of the SGS stress prediction is shown in Figure 37; the DS-ResNet has the highest accuracy of predictions for both dataset training and the testing one.

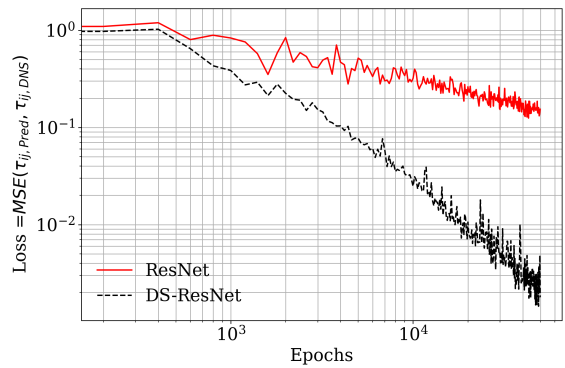

**Figure 33:** Mean Square Error (MSE) of the Sub Grid Scale (SGS) stress $\tau_{ij}$ between the DNS and the predictions by ResNet and DS-ResNet architectures through the training epochs, DS-ResNet architecture shows a dramatic increase in the training data learning speed and accuracy compared to the ResNet.

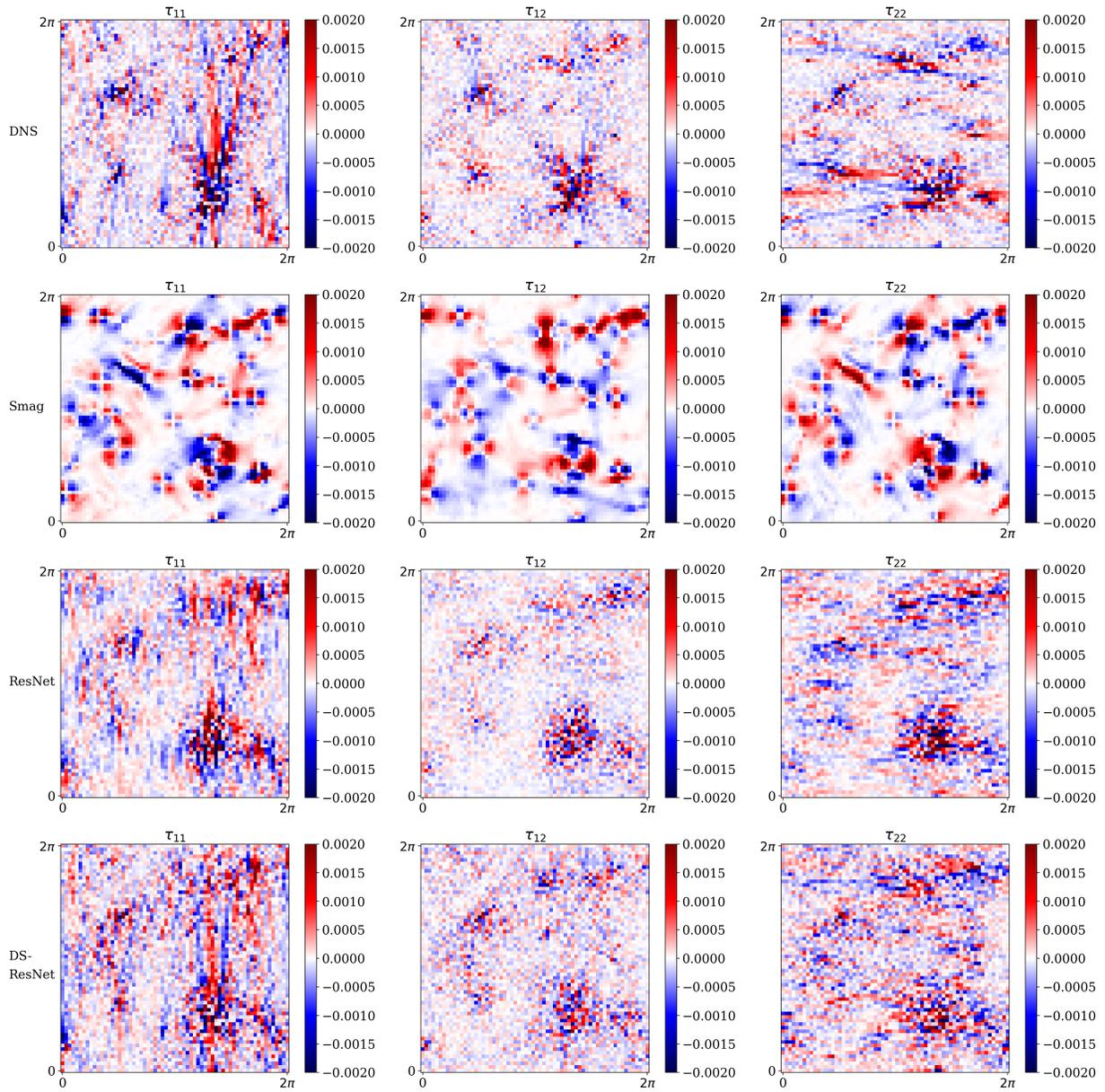

**Figure 34:** Sub Grid Scale (SGS) stresses $\tau_{ij}$ for the DNS, Smagorinsky model, ResNet, and the DS-ResNet for one of the frames in the testing dataset.

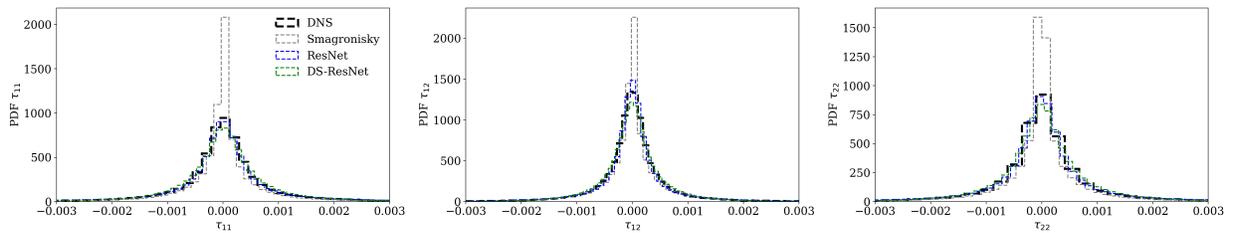

**Figure 35:** Probability Density Function (PDF) of the Sub Grid Scale (SGS) stresses $\tau_{ij}$ for the DNS, Smagorinsky model, ResNet, and the DS-ResNet for all testing dataset.

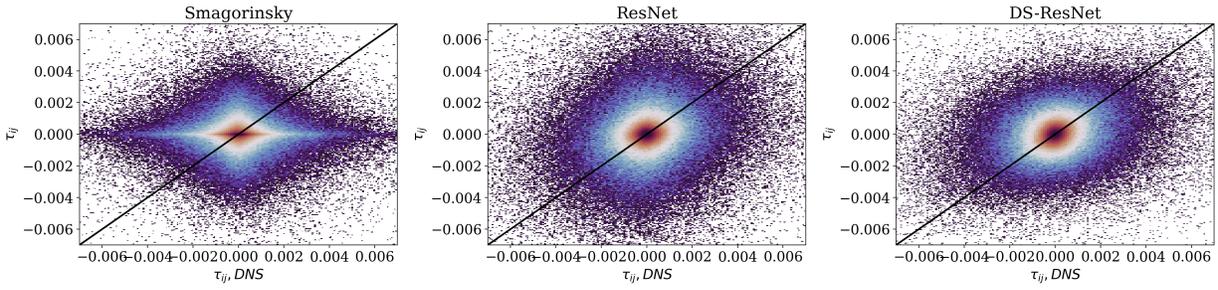

**Figure 36:** Joint Probability Density Function (PDF) of the Sub Grid Scale (SGS) stresses $\tau_{ij}$ for the DNS with the Smagorinsky model, ResNet, and the DS-ResNet for testing dataset.

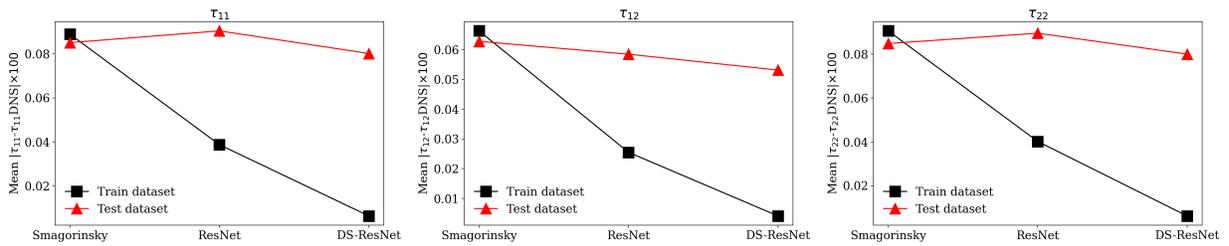

**Figure 37:** Sub Grid Scale (SGS) stresses $\tau_{ij}$ Mean Absolute Error (MAE) between the DNS SGS stresses and the three models (Smagorinsky, ResNet, and DS-ResNet) for the training and testing datasets, DS-ResNet has the least error.

## 5. Conclusion

In this study, the generative Residual Networks (ResNet) are utilized as a surrogate Machine Learning (ML) tool for Large Eddy Simulation (LES) Sub Grid Scale (SGS) resolving. In addition, the effect of using a Dual Scale Residual Block (DS-RB) in the ResNet architecture is studied. Two LES SGS resolving models are presented, a super-resolution model (SR-ResNet), and a SGS stress tensor inference one (SGS-ResNet). The SR-ResNet model task is to upscale a LES solution from coarse grids to finer ones by inferring the unresolved SGS velocity fluctuations. The SR-ResNet model is supervised, trained, and then tested with prior analysis. For example, low-resolution input data are derived from filtering and downsampling the high-resolution data. The model is trained on a 2-D planar LES single jet injection case at Reynolds number $1 \times 10^4$, and then tested on 2-D planar LES double jet injection cases and different Reynolds numbers at two different upscale ratios, 4 and 8. SR-ResNet showed success in upscaling the testing dataset, preserving the high-frequency velocity fluctuation information, and aligning with the energy spectrum of the higher-resolution LES solution. The results also show that using the Dual Scale Residual Blocks (DS-RB) in the SR-ResNet improves the prediction accuracy and precision of the high-frequency velocity fields compared to the Single Scale one (SS-RB), the results are presented in both spatial and spectral domains. For the SGS stress tensor inference (SGS-ResNet), the model is supervised trained, and tested on a filtered Direct Numerical Simulation (DNS) 2-D isotropic homogeneous turbulence problem dataset. Both SS-RB and DS-RB show a better prediction accuracy for the SGS stress tensor compared to the Smagoronisky model with reference to the true DNS SGS stress tensor, however, the DS-RB based SGS-ResNet is more statistically aligned with the DNS data. For both models, SR-ResNet and SGS-ResNet, the accuracy improvement while using DS-RB rather than the SS-RB is at the expense of network size, forward inference time, and training time. The increase in the network size is dramatic, more than 10 times. The increase in the training time and forward inference time is in order of about 0.5 and 3 times, respectively.

## 6. Future work

The present work only considered a prior analysis, by training and testing the models on filtered/downsampled LES and DNS datasets for both models, the super-resolution (SR-ResNet) model and the SGS stress tensor inference one (SGS-ResNet). Prior analysis for the present study is enough to show the proof of concept of using the ResNet and study the effect of the Dual Scale Residual Block (DS-RB). However, to test these models for more practical applications, the testing has to be implemented on a real LES dataset with no explicit filtration or downsampling. Implementing a posterior SR-ResNet for an unfiltered LES dataset or online/real-time SGS-ResNet LES stress tensor inference model requires extra model improvement to constrain the network's predictions for physically acceptable results and avoid model overfitting. The future model's improvements and implementation are summarized below:

- Include the pressure field data in the super-resolution and the SGS stress tensor inference task by physically constraining the model to satisfy the pressure-velocity incompressible Poisson equation.

- Implement transfer learning for the SR-ResNet model. This can be achieved by appending extra convolution layers to the SR-ResNet for only the inference. The original trained SR-ResNet layers will be frozen during the inference, while the appended layers will be trainable during the inference task to converge to a physically accepted solution, preventing overfitting.

- Incorporate unsupervised learning. This will enhance the models' generalization as it will help to train on versatile flow problems where no paired input-output datasets exist.

**Code and data availability**

The code and the data required to reproduce the presented results are publicly available on this GitHub repository: https://github.com/okhsallam/Dual-Scale-Residual-Network-for-2D-turbulent-flows/tree/main

**Declaration of competing interest**

The authors declare that they have no known competing financial interests or personal relationships that could have appeared to influence the work reported in this paper.